\documentclass[sn-basic]{sn-jnl}

\usepackage{graphicx}%
\usepackage{multirow}%
\usepackage{amsmath,amssymb,amsfonts}%
\usepackage{amsthm}%
\usepackage{mathrsfs}%
\usepackage[title]{appendix}
\usepackage{chngcntr}
\usepackage{xcolor}%
\usepackage{textcomp}%
\usepackage{manyfoot}%
\usepackage{booktabs}%
\usepackage{algorithm}%
\usepackage{algorithmicx}%
\usepackage{algpseudocode}%
\usepackage{listings}%
\usepackage{tabularx}
\usepackage{verbatim}

\usepackage{natbib}
\usepackage{apalike}
\usepackage{xpatch}

\usepackage{hyperref}
\hypersetup{
    colorlinks=true,
    citecolor=blue
}

\makeatletter
\xpatchcmd\NAT@citex
 {%
  \@citea\NAT@hyper@{%
    \NAT@nmfmt{\NAT@nm}%
    \hyper@natlinkbreak{\NAT@aysep\NAT@spacechar}{\@citeb\@extra@b@citeb}%
    \NAT@date
  }%
 }
 {%
  \@citea
  \NAT@nmfmt{\NAT@nm}%
  \NAT@aysep\NAT@spacechar
  \NAT@hyper@{\NAT@date}%
 }
 {}{}
\xpatchcmd\NAT@citex
 {%
  \@citea\NAT@hyper@{%
    \NAT@nmfmt{\NAT@nm}%
    \hyper@natlinkbreak{\NAT@spacechar\NAT@@open\if*#1*\else#1\NAT@spacechar\fi}%
    {\@citeb\@extra@b@citeb}%
    \NAT@date
  }%
 }
 {
  \@citea
    \NAT@nmfmt{\NAT@nm}%
    \NAT@spacechar\NAT@@open\if*#1*\else#1\NAT@spacechar\fi
    \NAT@hyper@{\NAT@date}%
 }
 {}{}
\makeatother

\begin{document}

\title[Article Title]{Novel second-order model for tumor evolution: description of cytostatic and cytotoxic effects}


\author*[1]{\fnm{Carlos~M.} \sur{Nieto}}\email{caniegue@correo.uis.edu.co}

\author*[1]{\fnm{Oscar~M.} \sur{Pimentel}}\email{oscarpimenteld214@gmail.com}

\author*[1]{\fnm{Fabio~D.} \sur{Lora-Clavijo}}\email{fadulora@uis.edu.co}

\affil*[1]{\orgdiv{Escuela de Física}, \orgname{Universidad Industrial de Santander}, \orgaddress{\street{Carrera 27 Calle 9}, \city{Bucaramanga}, \postcode{680002}, \state{Santander}, \country{Colombia}}}


\abstract{Cancer is a disease that takes millions of lives every year. Then, to propose treatments, avoid recurrence, and improve the patient's life quality, we need to analyze this disease from a biophysical perspective with a solid mathematical formulation. In this paper we introduce a novel deterministic model for the evolution of tumors under several conditions (untreated tumors and treated tumors using chemotherapy). Our model is characterized by a second-order differential equation, whose origin and interpretation are presented by exploiting our understanding of fluid mechanics (via continuity equations) and the theory of differential equations. Additionally, we show that our model can fit various experimental data sets. Thus, we prove that our nuanced and general model can describe accelerated growth, as well as cytostatic and cytotoxic effects. All in all, our model opens up a new window in the understanding of tumor evolution and represents a promising connection between the macroscopic and microscopic descriptions of cancer.
}

\keywords{Tumor evolution, cytostatic effects, cytotoxic effects, second-order model, Sturm-Liouville problem, continuity equation}

\maketitle

\section{Introduction}\label{sec1}

Cancer is a disease that has been affecting a large number of humans around the world. According to the World Health Organization, ‘in 2022, there were an estimated 20 million new cancer cases and 9.7 million deaths’ \citep{WHO}. This alarming number of cases has to be complemented by the number of patients whose access to health services, quality of life during the treatment, and the possibility of cancer recurrence. Therefore, more resources should be directed to the understanding of this disease. Given the complexity of cancer, arriving at full control of the disease has been a challenge \citep{WHO}. Studies on the dynamics of cancer span over a large number of directions. We believe that a deep understanding of cancer requires the collective work of medical doctors, patients, biologists, chemists, physicists, and mathematicians. Therefore, it is crucial to involve all areas of science to tackle this severe disease. In particular, a mathematical description involving physical laws, and biological and chemical processes, is needed to predict the evolution of a tumor. This description would allow physicians to determine the best treatment for a patient depending on the health state of the patient, the location of the tumor, and the time of diagnosis. It has been found that the evolution of a tumor entails a crucial number of biological processes. Among them, we have genetic mutations, deactivation of immune responses and growth suppressors, angiogenesis, and resistance to cell death \citep{Hallmarks}.\\

 We naturally expect the full description of cancer to be extremely complex. Its description requires concepts from molecular biophysics, fluid mechanics, and statistical mechanics. Several studies have been performed in this direction \citep{Cotner,DataDriven}. However, another approach has also been taken in the last decades \citep{simpleode,benzekry2014classical}. Given the nature of experiments regarding tumor evolution, several studies report the evolution of the size of a tumor in time. The information regarding the volume of a tumor can be seen as the macroscopic result of all the processes involved inside a tumor and its microenvironment. Therefore, tumor evolution can be understood as a population problem where cancer cells, normal cells, the immune system, and chemotherapy are at play. That approach has been used over the last decades and it has shown to be suitable in several scenarios. Those approaches are called volumetric models and they involve an ordinary differential equation of a variable $V$ (volume) as a function of time. The simplest model corresponds to the Exponential model, in which the volume grows without control up to large values (mathematically, it goes to infinity). Other common models are the Logistic model, the Gompertz model, and the Von Bertalanffy \citep{benzekry2014classical}. However, more complex models have been proposed over the last years (see, for instance, the work of \cite{WestNewton}). It is important to stress that tumor growth models play a crucial role in understanding the complex dynamics of cancer progression and guiding therapeutic interventions. By simulating tumor growth, researchers can gain valuable insights into the underlying mechanisms driving cancer development and assess the effectiveness of potential treatments. The integration of experimental data with these models allows for personalized predictions and represents a promising direction in understanding cancer as a whole. \\

Having a close look at the models discussed in the literature, we encountered two recurrent aspects: the models are derived from a first-order differential equation, and these models are used to describe growth and quasi-static evolution. It would be fruitful to have a model that can describe growth, quasi-stationary evolution, and shrinkage. It is possible to achieve such a goal by using a piecewise function containing the standard models and other simple functions (such as linear functions). However, having a continuous function would be more meaningful since it can give us information about the physics of the system. Thus,  we can get a hint about the origin of such macroscopic evolution of a tumor. On the other hand, having a higher-order differential equation would allow for a more nuanced evolution. \\

In this paper we propose a model that can describe tumor evolution under several conditions. Our model has the advantage of being adapted to several experimental data sets related to drug treatments. Namely, our model can account for usual growth, as well as cytostatic effects and cytotoxic effects.  To give strong arguments validating our model, we organize the paper as follows. In section \ref{sec:clamod} we discuss the most commonly known models used to describe tumor evolution, in section \ref{sec:secordmod} we introduce our general model and discuss how it can be used to fit several experimental data sets containing a variety of patterns.  In section \ref{sec:biophysicsexp} we discuss the possible origin and biophysical interpretation of our model; here, we talk about the interpretation of the parameters present in our model and their possible connection to medical parameters. We also argue why the second-order model nature of our model is expected from physical arguments and we exploit the mathematical properties of our formulation. Finally, we close with the Methods in section \ref{sec:methods} and the Conclusions in section \ref{sec:Con}. Additionally, we add two important appendices to understand the power of our model. In Appendix \ref{secA1} we discuss an interesting submodel that follows from our general formulation. In Appendix \ref{secA2} we show the fits for all the $85$ data sets analyzed in our study.\\

To make our discussion understandable, we clarify the nomenclature used in two of the references used to extract the experimental data points. We start with the data sets found in \cite{Ubezio}. In that article Figure 4 is crucial. There are two plots in that Figure: A and B. Additionally, each plot contains four curves: control, r (red), g (green), and v (violet). The case \emph{Control} corresponds to an untreated tumor, while the other data points are obtained using drug treatments. On the other hand, in \cite{falcetta2017modeling}
some abbreviations are worth clarifying. In that reference, three different drugs were employed:

\begin{itemize}
    \item DDP: Cisplatin
    \item BEV: Bevacizumab
    \item PTX: Paclitaxel
\end{itemize}

\noindent However, different particular doses and combinations were used in the study. We report the doses used in \cite{falcetta2017modeling} with the nomenclature employed in their plots and our plots:
\begin{itemize}
\item Control: No drug treatment whatsoever.
    \item DDP: Cisplatin, 3 mg/Kg every 	3 weeks.
\item BEV: Bevacizumab, 5 mg/Kg every e week.
\item PTXConv: Paclitaxel, 20 mg/Kg every three weeks.
\item PTX Equi: Paclitaxel, 8 mg/Kg weekly.
\item PTX High: Paclitaxel, 12 mg/Kg weekly.
\item DDP/PTX Conv: 3 mg/Kg of Cisplatin with 20 mg/Kg of PTX, both once every three weeks.
\item DDP/PTX Equi: 3 mg/Kg of Cisplatin once every three weeks with 8 mg/Kg of PTX weekly.
\item DDP/PTX High: 3 mg/Kg of Cisplatin once every three weeks with 12 mg/Kg of PTX weekly.
\item BEV/DDP/PTX Conv: 3 mg/Kg of Cisplatin and 5mg/Kg of Bevacizumab with 20 mg/Kg of PTX, all once every three weeks.
\item BEV/DDP/PTX Equi: 3 mg/Kg of Cisplatin and 5mg/Kg of Bevacizumab once every three weeks, with 8 mg/Kg of PTX weekly.
\item BEV/DDP/PTX High: 3 mg/Kg of Cisplatin and 5mg/Kg of Bevacizumab once every three weeks, with 12 mg/Kg of PTX weekly.

\end{itemize}

\section{Classical models for tumor growth} \label{sec:clamod}

In this section we discuss the most widely used classical models for tumor evolution, starting from the simplest to the most complex model.  The models presented in this section are characterized by first-order differential equations. It is important to highlight this property since our work presents arguments to go beyond this assumption. The motivation for having first-order models relies on their simplicity of interpretation. First-order differential equations are certainly easy to understand in terms of increasing and decreasing functions. However, this fact does not forbid other approaches to the problem of tumor evolution.

\subsection{Exponential model}

The Exponential model is one of the fundamental classical models for understanding tumor evolution. It assumes that tumor growth follows an exponential pattern, where the growth rate is directly proportional to the size of the tumor at a given time. This model can be used to describe uncontrolled growth and is characterized by the simple first-order differential equation
\begin{equation}
    \frac{dV}{dt}=\alpha V,
    \label{eq:class}
\end{equation}
whose solution for $V(t)$ takes the form
\begin{equation}
    V(t)=V_{0}e^{\alpha t},
    \label{eq:derexpmodel}
\end{equation}
where $V_{0}$ is the tumor volume at $t=0$ and
$\alpha$ characterizes the growth of $V(t)$. The parameter $\alpha$ can be understood as the initial fraction rate of change in $V$. Some authors call it the \emph{birth rate} \citep{Laleh}. However, this statement is true as long as $\alpha>0$. If the parameter $\alpha$ is negative, we have a decreasing evolution. It is worth mentioning that this model is also recognized for its limitations, as it may not fully capture the complexities of tumor growth dynamics, such as interactions with the tumor micro-environment or the development of resistance to treatment. Despite these limitations, the exponential model serves as a valuable starting point for studying tumor evolution.

\subsection{Logistic model}

The Logistic model is another classical model commonly used to describe tumor growth dynamics.  It assumes that tumor growth initially follows an exponential pattern, which eventually slows down leading to a plateau. This late-time behavior is often interpreted as a competition among proliferating tumor cells for space or nutrients \citep{vaghi2020population}. The first-order differential equation of this model contains the previous factor $\alpha V$ but it has an additional term with a negative sign that creates a transition towards a horizontal asymptote \citep{Verhulst,benzekry2014classical,Evain,Laleh}. The differential equation characterizing this model is usually written as
\begin{equation}
    \frac{dV}{dt}=\alpha V\left(1-\frac{V}{K}\right),
    \label{eq:logisticde}
\end{equation}
where $\alpha$ and $K$ are two positive constants. We see that the rate of change has a positive contribution (leading to growth) weighted by $\alpha$ and a negative contribution (leading to a decreasing behavior) weighted by $1/K$. The solution of this differential equation takes the form
\begin{equation}
    V(t)=\frac{KV_{0}}{V_{0}+(K-V_{0})e^{-\alpha t}},
    \label{eq:logisticsol}
\end{equation}
where $V_{0}$ is the volume at $t=0$. It is clear that, given the location of the exponential function, this expression goes asymptotically to a fixed value $V^{\infty}$ when $t\rightarrow\infty$. Taking the limit of (\ref{eq:logisticsol}) we get
\begin{equation}
V^{\infty}=\lim_{t\rightarrow\infty}V(t)=K.
\label{eq:VmaxLogistic}
\end{equation}
The constant $K$ corresponds to the carrying capacity of the tumor and represents the maximum and asymptotic value that the tumor volume can reach. It is important to note that this carrying capacity is a free parameter in the differential equation. In other words, it is not a derived quantity, contrary to what happens in our model discussed in the following sections. The logistic model offers a more realistic portrayal of tumor growth dynamics, albeit simplified compared to actual biological systems' complexities.

\subsection{Gompertz model}

The Gompertz model is another classical mathematical framework commonly utilized to describe tumor growth. Unlike the Exponential and Logistic models, the Gompertz model acknowledges the phenomenon of tumor growth deceleration over time. It was first introduced by \cite{Gompertz} and it can be presented in different ways. In this paper we adopt the convention given in \cite{vaghi2020population}. Therefore, the differential equation has the following form
\begin{equation}
    \frac{dV}{dt}=V\left[\alpha-\beta\ln\left(\frac{V}{V_{0}}\right)\right],
    \label{eq:gompertzde}
\end{equation}
where $\alpha$ and $\beta$ are two positive constants and $V_{0}$ is the initial tumor volume. We see that the presence of the parameter $\beta$ modifies the exponential model and it counteracts the uncontrolled growth driven by $\alpha$. The solution of (\ref{eq:gompertzde}) has the following form
\begin{equation}
    V(t)=V_{0}e^{\frac{\alpha}{\beta}(1-e^{-\beta t})},
    \label{eq:gompertzsol}
\end{equation}
where $V_{0}$ is the volume at $t=0$. This model describes a tumor that has an initial growth that ends in a horizontal asymptote at the value
\begin{equation}
V^{\infty}=\lim_{t\rightarrow\infty}V(t)=V_{0}e^{\frac{\alpha}{\beta}}.
\label{eq:VmaxGompertz}
\end{equation}
Therefore, we can say that there is also a carrying capacity whose value is $V_{0}e^{\frac{\alpha}{\beta}}$. The Gompertz model is characterized by a sigmoidal growth curve similar to the Logistic model but with more pronounced deceleration as the tumor approaches its maximum size. 

\subsection{Von Bertalanffy model}

We finish with a model that is also widely used in population growth studies. This model is also characterized by a first-order differential equation. This equation contains a positive term that leads to growth and a negative term that leads to a horizontal asymptote. This model is also a modification of the exponential model. However, there is a change in the sign of the linear term and the positive term is proportional to $V^{2/3}$. It is said that this last term is related to the area of the tumor and it leads to growth. The conventional form of the differential equation for the Von Bertalanffy model is \citep{VonBertalanffy,benzekry2014classical,Evain,Laleh} 
\begin{equation}
    \frac{dV}{dt}=\alpha V^{2/3}-\beta V,
    \label{eq:VonBerde}
\end{equation}
where $\alpha$ and $\beta$ are two positive constants. The solution of this differential equation can be written as
\begin{equation}
V(t)=\left[\frac{\alpha}{\beta}+\left(V_{0}^{1/3}-\frac{\alpha}{\beta}\right)e^{-\frac{\beta}{3}t}\right]^{3},
    \label{eq:VonBersol}
\end{equation}
where $V_{0}$ is the tumor volume at $t=0$. This model also exhibits initial growth and a transition towards a horizontal asymptote. Therefore, there is also a carrying capacity whose expression is given by
\begin{equation}
V^{\infty}=\lim_{t\rightarrow\infty}V(t)=\left(\frac{\alpha}{\beta}\right)^{3}.
\label{eq:VmaxVonBer}
\end{equation}
It is worth noticing that this maximum value does not depend on the initial value of the tumor volume. $V^{\infty}$ depends only on the parameters $\alpha$ and $\beta$. The Von Bertalanffy model postulates that growth is not only influenced by intrinsic factors but also by extrinsic environmental factors. \\

Other models in the literature are extensions of the previously discussed models \citep{Evain,Laleh}. Such models attempt to account for other properties, e.g., volume shrinkage. There are different ways of modifying the classical models. Some authors introduce new variables with the respective differential equation such that they get a system of coupled differential equations. This system of differential equations usually represents the interaction between the evolution of cancer cells, the action of the immune system, and the effect of cancer therapy. Other models include time-dependent parameters in the differential equations \citep{Evain,Laleh}. For instance, the carrying capacity in (\ref{eq:logisticde}) can be taken as a time-dependent term. Additionally, it is possible to define piecewise functions that account for the different phases that can exist in tumor evolution (e.g., growth phase, stationary phase, and shrinkage phase). In the next section we introduce a new model that can account for different phases observed in experimental data. Our model 
comes from a second-order differential equation and
contains parameters and time-dependent functions that resemble those in the classical models. 

\section{Second-order model} \label{sec:secordmod}

\subsection{General model}\label{generalmodel}
In this section we introduce our general model for tumor evolution, aiming to describe a wide spectrum of experimental data sets. As discussed in the Introduction, most models describe the exponential growth period and a stationary phase. However, experimental studies have shown that cytostatic and cytotoxic effects can also manifest in patients undergoing cancer treatments (e.g., chemotherapy and radiotherapy). Cytostatic effects are associated with a stationary phase or near-stationary phase, as observed in the traditional Logistic and Gompertz models. However, physicians primarily seek cytotoxic effects to shrink tumors. Therefore, we also require a model that accounts for tumor shrinkage. Moreover, other treatments have shown tumor recurrence after a stationary or shrinkage period. Hence, we need a new model that offers a wider spectrum of tumor dynamics and evolution. 

To describe a large number of experimental data regarding tumor growth evolution, we propose the following function for the volume as a function of time
\begin{equation}
    V=\frac{1}{Ae^{-a\phi(t)}+Be^{b\phi(t)}},
    \label{eq:mostgenV}
\end{equation}
where $a$ and $b$ are two positive constants, $A$ and $B$ are two integration constants and $\phi(t)$ is a function of time. The presence of the integration constants can be understood by introducing the differential equation leading to the function in (\ref{eq:mostgenV}). The two exponential functions in the denominator of the expression for $V(t)$ are associated with the increasing, stationary, and decreasing behaviors of the tumor volume as long as the parameters $a$ and $b$ are always positive and the function $\phi(t)$ is increasing, as will be clarified in the next sections. Equation (\ref{eq:mostgenV}) can be seen as the solution of the second-order differential equation
\begin{equation}
\frac{d^{2}V}{dt^{2}}-\frac{2}{V}\left(\frac{dV}{dt}\right)^{2}+\left[ \dot{\phi}(a-b)-\frac{\ddot{\phi}}{\dot{\phi}}\right]\frac{dV}{dt}+(\dot{\phi})^{2}abV=0.
\label{eq:mostgendeV}
\end{equation}
The second-order nature of this model makes it different from the traditional models discussed in the literature. The consequences of this fact will be discussed in the following sections. On the other hand, we can treat $V$ as a function of $\phi(t)$ (whose nature will be clarified soon). The differential equation of our model becomes
\begin{equation}
\frac{d^{2}V}{d\phi^{2}}-\frac{2}{V}\left(\frac{dV}{d\phi}\right)^{2}+(a-b)\frac{dV}{d\phi}+abV=0.
\label{eq:Vandphi}
\end{equation}
Since we have a second-order differential equation, the constants $A$ and $B$ are determined using the initial conditions $V(t=t_{0})=V_{0}$ and $\dot{V}(t=t_{0})=\dot{V}_{0}$. That is, we need to know the initial volume and its rate of change to derive the future behavior of the volume. The expressions for the integration constants $A$ and $B$ are
\begin{equation}
A=\frac{e^{a\phi(t_{0})}}{(a+b)V_{0}}\left[b+\frac{\dot{V}_{0}}{V_{0}\dot{\phi}(t_{0})}\right],
    \label{eq:genconstA}
\end{equation}
\begin{equation}
B=\frac{e^{-b\phi(t_{0})}}{(a+b)V_{0}}\left[a-\frac{\dot{V}_{0}}{V_{0}\dot{\phi}(t_{0})}\right].
    \label{eq:genconstB}
\end{equation}
To show the power of our model, we analyze particular cases in detail. The choice of these cases was made to describe all the experimental data found in \cite{falcetta2017modeling} and \cite{Ubezio}.

\subsection{Particular models}\label{subsec:particularmodels}
In this subsection we discuss the applicability of specific models derived from (\ref{eq:mostgenV}). These models are effective in describing tumor evolution under various medical conditions. 
However, before presenting all the selected particular models in our study, we introduce a form of our model for a specific choice of $\phi(t)$ that will be relevant throughout the paper. When we take a power-law form for this function $\phi(t)$, we get
\begin{equation}
    V=\frac{1}{Ae^{-a(t-c)^{n}}+Be^{b(t-c)^{n}}},
    \label{eq:polgenV}
\end{equation}
where $n$ is an integer number and $c$ is any real number. The value of $n$ will determine, for instance, how fast the volume increases or decreases, and the possible existence of a near-stationary phase. On the other hand, $c$ will determine the transition between phases as we will see in the next sections and subsections. For practical reasons, we introduce a notation that will be useful in our discussion. That is, we give a name to this choice for the function $\phi$ by defining
\begin{equation}
 \phi_{n}=(t-c)^{n}.
\end{equation}
When we use this form for the function $\phi(t)$ in the expressions for the constants $A$ and $B$ of Eqs. (\ref{eq:genconstA}-\ref{eq:genconstB}), we get contributions to the exponential function in Eq. (\ref{eq:polgenV}). For instance, when $t_{0}=0$, we obtain a modified factor that depends on $c$ within the exponential functions in (\ref{eq:polgenV}). Therefore, it is suitable to define 
\begin{equation}
 \tilde{\phi}_{n}=\left[(t-c)^{n}-(-c)^{n}\right].
 \label{eq:newphi}
\end{equation}
Now we define another quantity that will be important in this work: the relative volume at a time $t$ concerning the initial volume $V_{0}=V(t_{0})$ as follows
\begin{equation}
V_{R}(t)=\frac{V(t)}{V_{0}}.
\label{eq:relvol}
\end{equation}
In general, the resulting behavior of $V$ depends on the choices of the parameters, the initial conditions, and the function $\phi$. Therefore, we split the particular models into two big different sets: models with terminal shrinkage and models without terminal shrinkage. Since one of our goals is to understand cytostatic and cytotoxic effects from a quantitative perspective, this two-set separation allows us to highlight the most relevant aspects of each particular model. The two sets are characterized by the value of the parameter $b$. If $b=0$, we obtain models without terminal shrinkage. If $b\neq0$, we obtain models with terminal shrinkage. \\

In this section we demonstrate the power of this separation and its application to fit experimental data. In the following sections we explore the biophysical implications of having  $b=0$ or $b\neq0$. We refer to the set with $b=0$ as \emph{models with terminal shrinkage} because they all satisfy the following property
\begin{equation}
    \lim_{t\rightarrow\infty}V_{R}(t)\rightarrow0.
    \label{eq:remvrlimit}
\end{equation}
In other words, the relative volume goes to zero at large times. This result implies that the final volume will be much smaller than the initial volume $V_{0}$. Essentially, there is no recurrence at that particular location in the body. The phenomenon of recurrence or regrowth will be observed in some models in the other set: \emph{models without terminal shrinkage}. The models in this big set do not have a $V$ approaching zero for large $t$. Instead, they either exhibit a divergent behavior
\begin{equation}
    \lim_{t\rightarrow\infty}V_{R}(t)\rightarrow \infty,
    \label{eq:noremvrlimitI}
\end{equation}
or the volume reaches a finite maximum value $V_{R}^{\infty}$
 \begin{equation}
    \lim_{t\rightarrow\infty}V_{R}(t)\rightarrow V_{R}^{\infty}.
    \label{eq:noremvrlimitII}
\end{equation}

\subsubsection{Models without terminal shrinkage}

Before starting the detailed discussion of the innovative models derived from Eq. (\ref{eq:polgenV}), we stress that our general Eq. (\ref{eq:mostgenV}) contains the traditional exponential model, the Gompertz model and a modification of the Logistic model. The \emph{exponential model} is obtained with the following choice: $\phi(t)=t$, $b=0$ and $a=\dot{V}_{R0}$. The \emph{Gompertz model} is obtained with the following choice: $b=0$, $\phi(t)=-e^{-\beta t}$ and $a=\dot{V}_{R0}/\beta$ (with $\beta$ a positive constant). On the other hand, the model that represents a \emph{modification of the Logistic model} is obtained with the choice: $b=0$ and $\phi(t)=t$. In the following lines, we discuss in detail some properties of this \emph{modified logistic model} and compare it with two traditional models.\\

\noindent {\bf Modified logistic model}\\

\begin{figure}[H]
\centering
\includegraphics[width=\textwidth]
{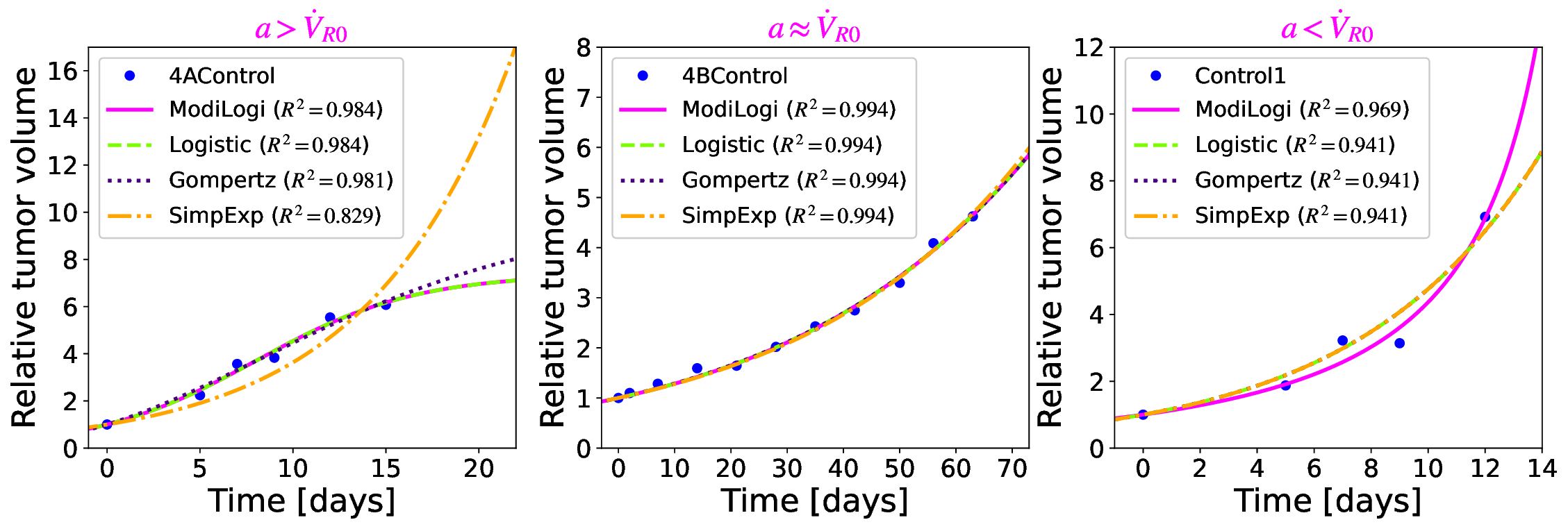}
\caption{Comparison of the models shown in Eq. (\ref{eq:fourmodels}) using data from \cite{Ubezio} in the left and middle panels, and data from \cite{falcetta2017modeling} in the right panel. The dots represent the experimental data points and the lines represent the fits as explained in each legend. On top of each panel we wrote in magenta the relation between $a$ and $\dot{V}_{R0}$ predicted by our modified logistic model. Moreover, we used the label \emph{ModiLogi} for our modified logistic model and \emph{SimpExp} for the simple/traditional exponential model}
\label{Fig:Comparison}
\end{figure}

We refer to this case as the \emph{modified logistic model} because it shares a similar structure with the traditional logistic model in Eq. (\ref{eq:logisticsol}). Although the traditional and the modified logistic models have some similarities, they are not equivalent. The explicit expression of our modified logistic model is
\begin{equation}
    V_{R}(t)=\frac{a}{\left(\dot{V}_{R0}\right)e^{-at}+\left(a-\dot{V}_{R0}\right)},
    \label{eq:newlogistic}
\end{equation}
where we have set $t_{0}=0$. It is important to note the presence of $(a-\dot{V}_{R0})$ in the denominator. The value and sign of this term define the overall behavior of the function $V_{R}(t)$. In the following lines, we want to show the properties of our modified logistic model by discussing three distinct cases and by comparing this new model with three widely used models in the literature (exponential, logistic, and Gompertz). We rewrite here the equations for the relative volume in each formulation for completeness
\begin{align}
    &V_{R}^{E}(t)=e^{\alpha t}, \ \hspace{2.5cm} V_{R}^{G}(t)=e^{\frac{\alpha}{\beta}\left(1-e^{-\beta t}\right)}, \notag \\ 
    &V_{R}^{L}(t)=\frac{K_{R}}{1+\left(K_{R}-1\right)e^{-\alpha t}}, \ \ V_{R}^{ML}(t)=\frac{a}{\left(\dot{V}_{R0}\right)e^{-at}+\left(a-\dot{V}_{R0}\right)},
    \label{eq:fourmodels}
\end{align}
where $K_{R}=K/V_{0}$ is the relative carrying capacity and we have used the following labels for the models: simple exponential (E), Gompertz (G), logistic (L), modified logistic (ML). 
The simplest situation in the modified logistic model occurs when $a=\dot{V}_{R0}$. This relation leads to the usual exponential model, as explained before. We use the expression \emph{simple exponential} because we introduce extensions of this model in the following pages. In the middle panel of Fig. \ref{Fig:Comparison} we show the four models for one data set found in \cite{Ubezio}. In this case, our model predicts that $a\approx\dot{V}_{R0}$ and all the four models have similar behavior, which is an exponential evolution during the time period of interest. The relative carrying capacity predicted by the logistic and modified logistic models is of the order $50$, while the Gompertz model predicts a relative carrying capacity of the order $8\times10^{8}$. Thus, the evolution seen in all models is practically an exponential growth over the first 100 days of the evolution.  \\

In the case $a>\dot{V}_{R0}$, we get an evolution that is similar to the logistic and Gompertz models because there is an initial fast growth and then the volume reaches a plateau after some time. The maximum value reached by $V_{R}$ in our modified logistic model is
\begin{equation}
\lim_{t\rightarrow\infty}V_{R}=V_{R}^{\infty}=\frac{a}{a-\dot{V}_{R0}}.
\label{eq:maxvn1}
\end{equation}
This stationary region brings us the notion of carrying capacity. In the case at hand, the carrying capacity is not a parameter in the differential equation, it is rather a quantity depending on the initial growth rate $\dot{V}_{R0}$ and the parameter $a$. It is important to note that this carrying capacity can go to infinity when $a=\dot{V}_{R0}$. In such a case, we recover the exponential model. That is, an infinite carrying capacity corresponds to uncontrolled exponential growth. In the left panel of Fig. \ref{Fig:Comparison} we show the comparison of four fits using our modified logistic model, the logistic model, the Gompertz model, and the exponential model. We see that our modified logistic model, the logistic model, and the Gompertz model have all similar behavior. According to the value of $R^{2}$, our modified logistic model and the traditional logistic model show the best fit.  \\

On the other hand, the case $a<\dot{V}_{R0}$ is both novel and interesting. When $a<\dot{V}_{R0}$, a singularity arises that splits the function into two branches. Focusing on the interval going from $t_{0}=0$ to the asymptote, there is fast growth. The increase in $V_{R}$ occurs faster than in the exponential case. If we analyze (\ref{eq:maxvn1}), we find that the value of $V_{R}$ becomes negative when $t$ approaches infinity. This result might suggest a lack of validity of the model. However, we have to keep in mind that we are not interested in the branch that lives beyond the asymptote. That part of the function exists from a mathematical point of view, but it does not have any biophysical meaning. The expression associated to the vertical asymptote is given by
\begin{equation}
    t_{asymp}=\frac{1}{a}\ln\left[\frac{\dot{V}_{R0}}{\dot{V}_{R0}-a}\right].
    \label{eq:vertasymp}
\end{equation}
In the right panel of Fig. \ref{Fig:Comparison} we observe that the logistic, the Gompertz, and the exponential models have all a similar evolution. On the contrary, our modified logistic model stands out as the best description both visually and quantitatively. This $a<\dot{V}_{R0}$ case is a novel evolution that can be understood in terms of the relation between $a$ and $\dot{V}_{R0}$. As will be explained in more detail in the sub-section \ref{subsec:parammodels}, having a parameter $a$ that is smaller than the initial growth rate is insufficient to stop the growth and the tumor gets larger and larger with time.

From the form of Eq. (\ref{eq:newlogistic}) and the magenta fits in Fig. \ref{Fig:Comparison}, we can make some conclusions regarding the role of the parameters $a$ and $\dot{V}_{R0}$. As expected, $\dot{V}_{R0}$ determines how fast the volume will start increasing with time, while $a$ is related to the stationary transition. There is only a stationary region or plateau when $a>\dot{V}_{R0}$. That is, there is a cytostatic effect when $a$ is larger than the initial growth rate. If $a$ is smaller or equal to $\dot{V}_{R0}$, there is no stationary phase; that is, there is no carrying capacity and the volume increases to large values when $t$ is also large. However, there is a difference between these two cases. When $a$ is equal to $\dot{V}_{R0}$, there is a balance that is not enough to stop the growth of the tumor; this relation leads to traditional exponential growth. Instead, when $a$ is smaller than $\dot{V}_{R0}$, a vertical asymptote arises in the expression for the relative volume. This asymptote makes $V_{R}$ go quicker to very large values.

In the left panel of Fig. \ref{Fig:Comparison}, we see that our modified logistic model predicts a plateau and then a maximum value for $V_{R}$ 
that is a sign of cytostatic effects. Taking $t_{0}=0$, the value of $t$ associated with this transition from the growth phase to the stationary phase is equal to
\begin{equation}
    t_{tr}=\frac{1}{a}\ln\left[\frac{\dot{V}_{R0}}{a-\dot{V}_{R0}}\right].
    \label{eq:trantimemodilog}
\end{equation}
If the value of this time and the maximum relative volume are very high, the tumor will have harmful effects on the patient. Therefore, one of the goals of cancer treatment is to make $t_{tr}$ and $V_{R}^{\infty}$ as small as possible. A specific treatment might make $t_{tr}$ and $V_{R}^{\infty}$ smaller or they might produce a different behavior. The outcome will depend on the treatment and the living organism. From Fig. \ref{Fig:Comparison} we show that our modified logistic model is an alternative option to the traditional models. However, our original formulation goes beyond the traditional models as we explicitly explain in the following lines.\\

\noindent \emph{{\bf Generalized modified logistic model of degree $n$}}: \\

\begin{figure}[H]
\centering
\includegraphics[scale=0.6]{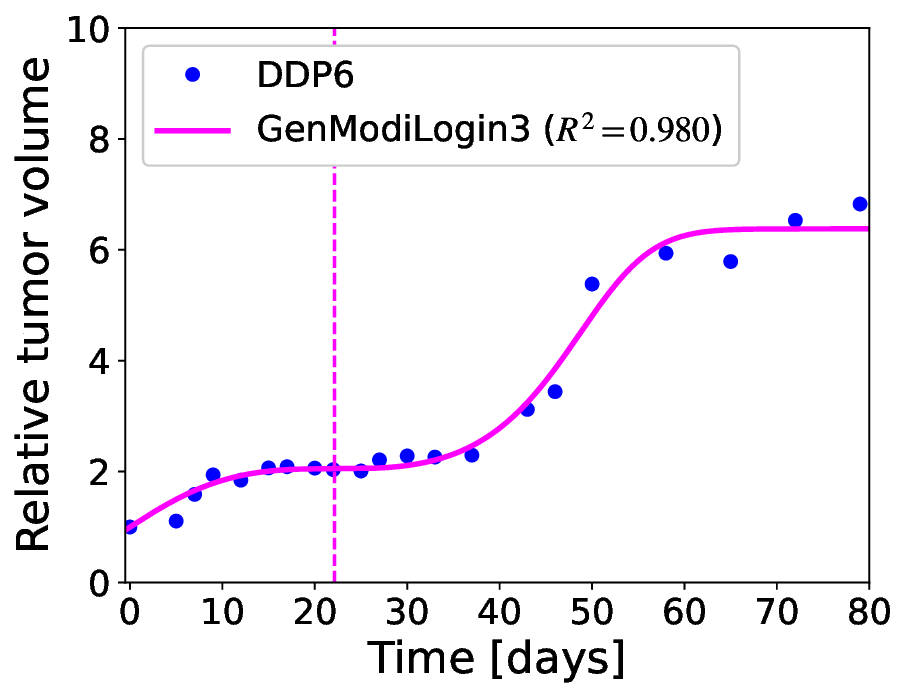}
\caption{Result of the fitting process for the data set \emph{DDP6} from \cite{falcetta2017modeling} using Eq. (\ref{eq:genmodlog3}): the dots represent the experimental data and the continuous line represents the best fit. The dashed vertical lines show the value of $c$}
\label{Fig:GenModLog}
\end{figure}

\noindent We now discuss the straightforward generalization of our modified logistic model. It is also obtained by taking $b=0$ in (\ref{eq:polgenV}). We call this function the \emph{generalized modified logistic model of degree $n$} because it has the same structure as (\ref{eq:newlogistic}) but it allows for different values of the parameter $n$. The explicit expression for this model is
\begin{equation}
   V_{R}(t)=\frac{a}{\left(\frac{\dot{V}_{R0}}{n(-c)^{n-1}}\right)e^{-a((t-c)^{n}-(-c)^{n})}+\left(a-\frac{\dot{V}_{R0}}{n(-c)^{n-1}}\right)},
    \label{eq:genmodlog}
\end{equation}
for $c\neq0$. As clear from our explicit form of the function, this solution is only valid when $c\neq0$. If $c=0$, we derive an alternative equation from (\ref{eq:polgenV}). Thus, when $c=0$, the correct expression to use is given by
\begin{equation}
   V_{R}(t)=\frac{1}{\tilde{A}e^{-at^{n}}+\left(1-\tilde{A}\right)},
    \label{eq:genmodlognoc}
\end{equation}
When $n=3$, we get the form
\begin{equation}
    V_{R}(t)=\frac{a}{\left(\frac{\dot{V}_{R0}}{3c^{2}}\right)e^{-a\left[(t-c)^{3}+(c)^{3}\right]}+\left(a-\frac{\dot{V}_{R0}}{3c^{2}}\right)},
        \label{eq:genmodlog3}
\end{equation}
for $c\neq0$, or
\begin{equation}
    V_{R}(t)=\frac{1}{Ae^{-at^{3}}+\left(1-\tilde{A}\right)},
        \label{eq:genmodlog3noc}
\end{equation}
for $c=0$. These models belong to the set of \emph{models without terminal shrinkage} because they never go to zero when $t\rightarrow\infty$. According to the value of $\Delta=\left(a-\frac{\dot{V}_{R0}}{3c^{2}}\right)$ in (\ref{eq:genmodlog3}), we can have three types of behavior. When $\Delta$ is equal to zero we get uncontrolled growth; this case will be analyzed in detail in the following pages and will be called \emph{generalized exponential model of degree $n$}. When $\Delta<0$, we get a vertical asymptote in the function; this behavior was not found in the analyzed experimental data. However, this case might be interesting in future studies. Finally, when $\Delta>0$ we get a function that resembles the logistic model in the sense that $V_{R}$ reaches a plateau when $t$ approaches infinity. However, in this case, there is interestingly an additional stationary region. In Fig \ref{Fig:GenModLog} we can see this function fitting an experimental data set. We observe that the function reaches a near-stationary region at some point. However, $V_{R}$ resumes its growth until it reaches a definite stationary region that leads to a maximum $V_{R}^{\infty}$ when $t\rightarrow\infty$. From a biophysical point of view, the first stationary region of our model corresponds to a cytostatic effect on the tumor that lasts for a short period. At some point, the volume increases again until it reaches a second and definite stationary region. It is interesting to see that the existence of two stationary regions is predicted by our generalized model and that there is experimental evidence of this effect. The general expression for the maximum volume with a general $n$ is  
\begin{equation}
   V_{R}^{\infty}=\frac{a}{\left(a-\frac{\dot{V}_{R0}}{n(-c)^{n-1}}\right)}.
    \label{eq:vmaxgenmodlog}
\end{equation}

\noindent The model that follows corresponds to a particular case of the previous model. However, it is important to discuss it in detail because it is connected to interesting interpretations in terms of treatment effects. We call it \emph{exponential} in the sense that it predicts a relative volume $V_{R}$ that goes to infinity when $t\rightarrow\infty$. The way it goes to infinity differs from the traditional exponential model but we observe uncontrolled growth that is associated with the inefficacy or absence of treatment at large times.\\

\noindent \emph{{\bf Generalized exponential model of degree $n$}}:\\

\noindent Now we go beyond the traditional exponential model. To generalize the exponential model, we start with the generalized modified logistic model of degree $n$ of Eq. (\ref{eq:genmodlog}) and take
\begin{equation}
    a=\frac{\dot{V}_{R0}}{\dot{\phi})t_{0}}=\frac{\dot{V}_{R0}}{n(t_{0}-c)^{n-1}}=\frac{\dot{V}_{R0}}{n(-c)^{n-1}},
\end{equation}
where we have assumed that $t_{0}=0$. Thus, the model takes the form
\begin{equation}
    V_{R}(t)=e^{\left(\frac{\dot{V}_{R0}}{n(-c)^{n-1}}\right)\left[(t-c)^{n}-(-c)^{n}\right]},
    \label{eq:genexpmodn}
\end{equation}
when $c\neq0$. If $c=0$, we have to derive a new equation starting from our original form (\ref{eq:polgenV}). We can also start from (\ref{eq:genmodlog3noc}) and take $\tilde{A}=1$. Thus, we arrive at
\begin{equation}
    V_{R}(t)=e^{at^{n}},
    \label{eq:genexpmodnnoc}
\end{equation}

\begin{figure}[H]
\centering
\includegraphics[scale=0.6]{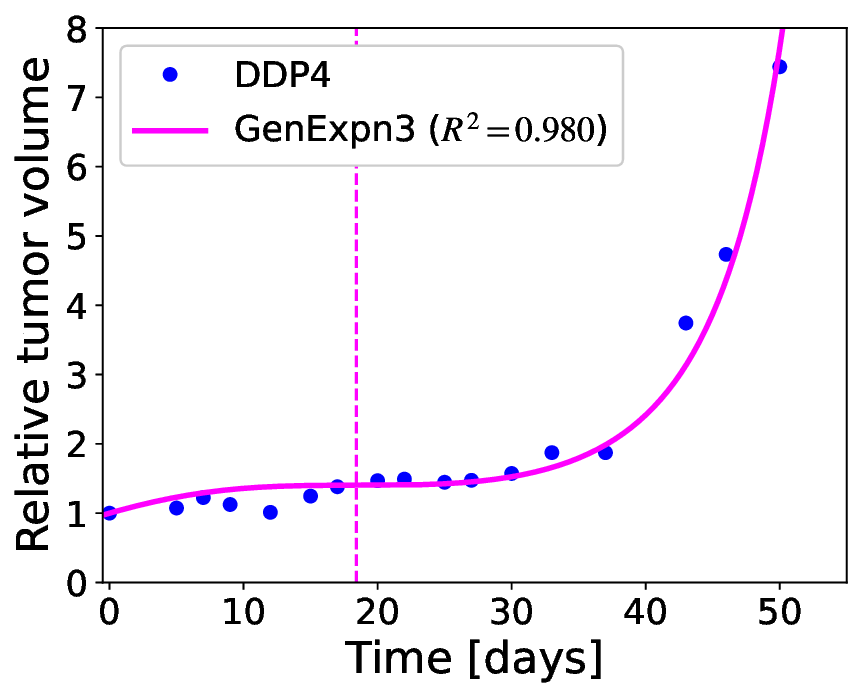}
\caption{Result of the fitting process for the data set \emph{DDP4} from \cite{falcetta2017modeling} using Eq. (\ref{eq:genexpmod3}): the dots represent the experimental data and the continuous line represents the best fit. The vertical line marks the position of $c$}
\label{Fig:GenExp}
\end{figure}

These generalized exponential models allow us to have an intermediate near-stationary period followed by fast growth. That is, they represent an extension of the traditional exponential model, where there is only fast growth. Several data sets from the sources used  \citep{falcetta2017modeling,Ubezio} can be fitted with this general model. All these fits are found with $n=2$ or $n=3$. There are some cases in which $c=0$, and other cases in which $c\neq0$. In the case of a non-zero $c$, we have the function
\begin{equation}
    V_{R}(t)=e^{\left(\frac{\dot{V}_{R0}}{3(c)^{2}}\right)\left[(t-c)^{3}+(c)^{3}\right]},
    \label{eq:genexpmod3}
\end{equation}
for $n=3$, and
\begin{equation}
    V_{R}(t)=e^{\left(\frac{-\dot{V}_{R0}}{2c}\right)\left[(t-c)^{2}-(c)^{2}\right]},
    \label{eq:genexpmod2}
\end{equation}
for $n=2$. Clearly, in this last case $\dot{V}_{R}(0)$ must be less than zero such that $a$ remains positive as defined since the beginning of our work.
In Fig. \ref{Fig:GenExp} we show one of the data sets fitted with the generalized exponential model with $n=3$ and $c\neq0$. There is a near-stationary region and fast growth afterward. The value of $c$ indicates the point where there is a change in the concavity of the function. This fact can be understood by noting that both the first and second derivatives of $V_{R}$ vanish at $t=c$. Different values of $n$ would modify the \emph{near-stationary} region. In Subsection \ref{subsec:parammodels} we discuss the role of $n$ in more detail. Additionally, in Appendix \ref{secA2} we show fitted data with different values of $n$ and $c$.\\

\subsubsection{Models with terminal shrinkage}

We now turn our attention to models exhibiting terminal volume shrinkage. All these models are characterized by non-zero parameters $a$ and $b$. As we did in the previous subsection, we start from the simplest possible model and then we move to more complex scenarios. The complexity of such models is linked to the rich dynamics involved in the evolution of a given tumor. According to our knowledge of chemotherapy, volume shrinkage is associated with a cytotoxic effect of drugs on the tumor. Thus, in this subsection we deal with models that can describe volume shrinkage due to cytotoxic drugs and the effect on the immune system.\\

\noindent \emph{{\bf Asymmetric shrinkage model of first degree}}:\\

\begin{figure}[H]
\centering
\includegraphics[scale=0.42]{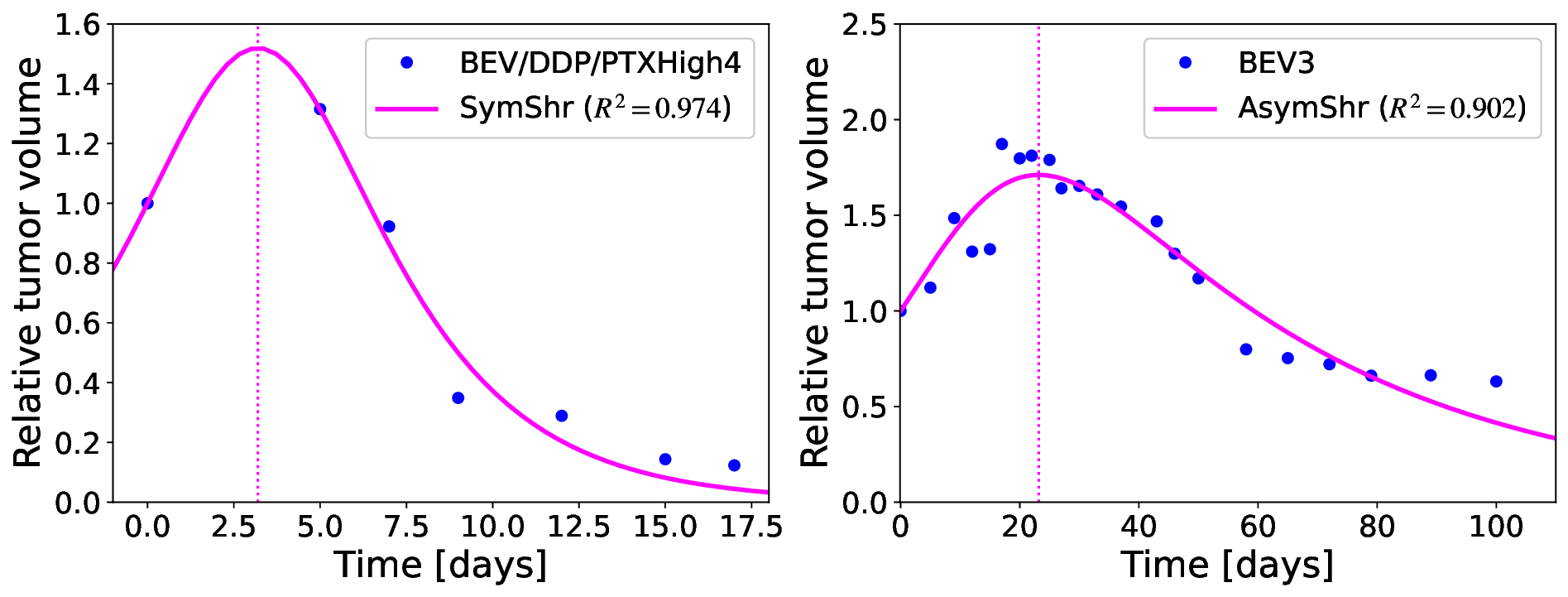}
\caption{Result of the fitting process: the dots represent the experimental data and the continuous line represents the best fit. {\bf Left:} Data set \emph{BEV/DDP/PTXHigh4} from \cite{falcetta2017modeling} fitted using Eq. (\ref{eq:shrinkageab}). {\bf Right:} Data set \emph{BEV1} from \cite{falcetta2017modeling} fitted using Eq. (\ref{eq:symmshr}). The dotted vertical lines represent the points where each function has its maximum value}
\label{Fig:RemMod}
\end{figure}

We start with the simplest model. That is, we start with the case in which $\phi(t)=t$. If we take Eq. (\ref{eq:polgenV}) and use this linear function, we arrive at
\begin{equation}
    V_{R}(t) = \frac{(a+b)}{(b+\dot{V}_{R0})e^{-at}+(a-\dot{V}_{R0})e^{bt}},
    \label{eq:shrinkageab}
\end{equation}
where we have taken $t_{0}=0$. If $a\neq b$, we obtain an asymmetric function. That is, the growth phase and the shrinkage phase have different behavior. On the other hand, given the increasing and decreasing regions of the function $V_{R}$, there is a maximum at
\begin{equation}
t_{cri}=\frac{1}{(a+b)}\ln\left(\frac{a(b+\dot{V}_{R0})}{b(a-\dot{V}_{R0})}\right).
\label{eq:maximumshrab}
\end{equation}
The position of the maximum depends on $a$, $b$, and the initial growth rate $\dot{V}_{R0}$. Most treatments aim at reaching this maximum in the smallest possible time. In medical terms, a smaller $t_{crit}$ generally implies a more effective treatment, although it is important to take side effects into account \citep{Markman,Ooki}. In the right panel of Fig. \ref{Fig:RemMod}, we show a data set that is fitted with this model. There is a growth phase and a shrinkage phase. \\

\noindent \emph{{\bf Symmetric shrinkage model of first degree}}:\\

If we take $a=b$ in (\ref{eq:shrinkageab}), we get the symmetric model
\begin{equation}
    V_{R}(t) = \frac{2a}{(a+\dot{V}_{R0})e^{-at}+(a-\dot{V}_{R0})e^{at}}.
    \label{eq:symmshr}
\end{equation}
This model is called symmetric because it presents a symmetry around the critical time
\begin{equation}
    t_{crit}=\frac{1}{2a}\ln(A/B)=\frac{1}{2a}\ln\left(\frac{a+\dot{V}_{0}/V_{0}}{a-\dot{V}_{0}/V_{0}}\right)=\frac{1}{2a}\ln\left(\frac{a+\dot{V}_{R0}}{a-\dot{V}_{R0}}\right).
    \label{eq:crittimesymm}
\end{equation}
In the left panel of Fig \ref{Fig:RemMod} we show a data set that is fitted using this symmetric model. We observe a mirror symmetry concerning the position of the maximum. The presence of such a symmetry might not be evident from (\ref{eq:symmshr}). However, we can show that this model presents a time-inversion symmetry. First, we introduce a new time variable $\tau$ such that 
\begin{equation}
    t=\tau+t_{crit},
\end{equation}
where $t_{crit}$ is the critical time given in (\ref{eq:crittimesymm}). Thus, we obtain the following expression for the relative volume
\begin{equation}
    V_{R}(\tau)=\frac{2a}{(a-\dot{V}_{R0})^{1/2}(a+\dot{V}_{R0})^{1/2}}\cdot\frac{1}{e^{-a\tau}+e^{a\tau}}
\end{equation}
From this result, it is clear that 
\begin{equation}
    V_{R}(-\tau)=V_{R}(\tau).
\end{equation}
We conclude then that the function $V_{R}$ is symmetric around the critical point (\ref{eq:crittimesymm}).\\

It is important to note that these first two models do not describe long-term cytostatic effects. That is, there is a sharp transition from growth to shrinkage. We argue that this fact is due to the linear function $\phi(t)$ that we chose. Therefore, we postulate that cytostatic effects can be present if we take a power-law form for $\phi(t)$. Thus, both cytostatic and cytotoxic effects can be obtained by keeping $b\neq0$ and choosing a function $\phi(t)=(t-c)^{n}.$ We discuss this case in the next lines.\\

\noindent \emph{{\bf Generalized shrinkage model of degree $n$}}:\\

\noindent We end our discussion of particular models with a function that contains both cytostatic and cytotoxic effects. As discussed above, we already know that a non-zero $b$ induces a terminal-shrinkage effect and a power-law $\phi$ induces a near-stationary phase. Therefore, we take the expression
\begin{equation}
    V_{R}(t)=\frac{(a+b)}{\left(b+\frac{\dot{V}_{R0}}{n(-c)^{n-1}}\right)e^{-a\tilde{\phi}_{n}}+\left(a-\frac{\dot{V}_{R0}}{n(-c)^{n-1}}\right)e^{b\tilde{\phi}_{n}}},
    \label{eq:genshr}
\end{equation}
where $\tilde{\phi}_{n}$ was defined in (\ref{eq:newphi}). It is important to stress that the equation is valid as long as $c\neq0$. If $c$ is equal to zero, we have to derive an expression that is valid in that case only. Starting from (\ref{eq:polgenV}), we obtain that the relative volume will depend on four numbers: $a$, $b$, $\tilde{A}$ and $n$
\begin{equation}
    V_{R}(t)=\frac{1}{\tilde{A}e^{-at^{n}}+\left(1-\tilde{A}\right)e^{bt^{n}}},
    \label{eq:genshrnoc}
\end{equation}
where the parameter $\tilde{A}$ will depend on the lowest non-zero derivative of $V_{R}$. That is, $\tilde{A}$ will depend on $V^{(n)}(0)$.\\

\begin{figure}[H]
\centering
\includegraphics[scale=0.5]{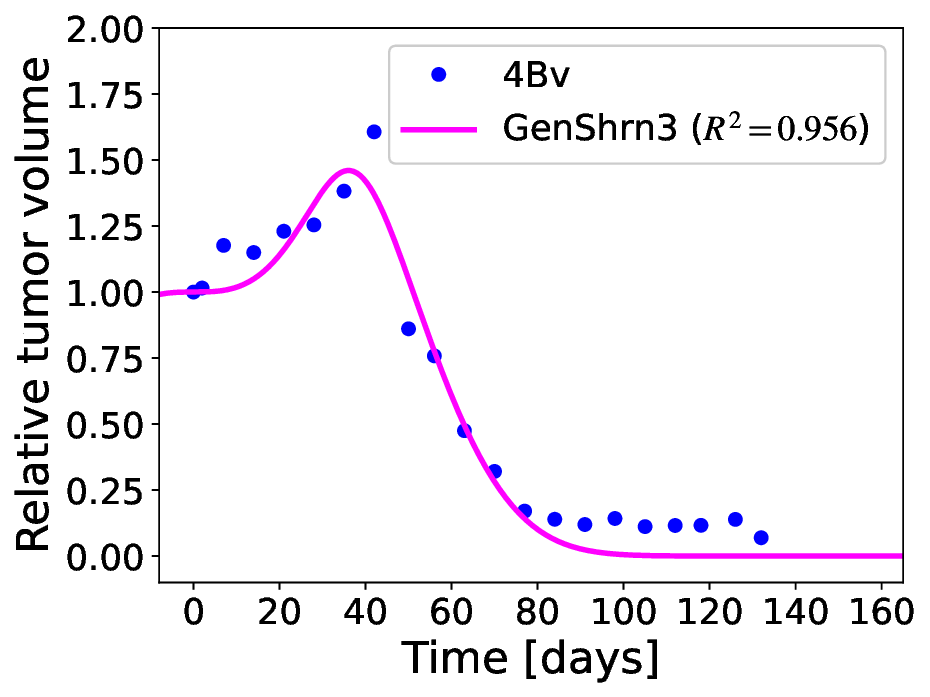}
\caption{Fit of the data set \emph{4B}v from \cite{Ubezio} using Eq. (\ref{eq:genshr3noc}). We show the fit in a larger domain to appreciate the form of the function}
\label{Fig:genshrn3vl}
\end{figure}

On the other hand, since there is a growth phase and a shrinkage phase, there should be at least a maximum value of $V_{R}$. Working with the general equation (\ref{eq:polgenV}), we get the following expression for the critical time
\begin{equation}
t_{cri}=\left[\frac{1}{(a+b)}\ln\left(\frac{a\left[b+\frac{\dot{V}_{R0}}{n(-c)^{n-1}}\right]e^{a(-c)^{n}}}{b\left[a-\frac{\dot{V}_{R0}}{n(-c)^{n-1}}\right]e^{-b(-c)^{n}}}\right)\right]^{\frac{1}{n}}+c.
\end{equation}
Depending on the value of $n$ we can have more than one maximum If $n$ is odd, we get only one $t_{crt}$; if $n$ is even, we get two values for $t_{cri}$. In our fitting process, we found that $n=3$ is an optimal value for one of the data sets in \cite{Ubezio}. Therefore, we use the following specific model
\begin{equation}
    V_{R}(t)=\frac{(a+b)}{\left(b+\frac{\dot{V}_{R0}}{3c^{2}}\right)e^{-a\tilde{\phi}_{3}(t)}+\left(a-\frac{\dot{V}_{R0}}{3c^{2}}\right)e^{b\tilde{\phi}_{3}(t)}},
    \label{eq:genshr3}
\end{equation}
for $c\neq0$, or 
\begin{equation}
    V_{R}(t)=\frac{1}{\tilde{A}e^{-at^{3}}+\left(1-\tilde{A}\right)e^{bt^{3}}},
    \label{eq:genshr3noc}
\end{equation}
for $c=0$
In Fig \ref{Fig:genshrn3vl} we see a data set that is fitted using this model with a zero $c$. The initial growth is less tilted compared to the previous models. In fact, the first derivative of $V_{R}$ is zero at $t_{0}=0$. This zero derivative is the result of the form (\ref{eq:genshr3noc}). Even though $\dot{V}_{R}(0)=0$, the volume increases for $t>0$ because the third derivative is not zero.
The behavior seen with this model contains a near-stationary period, a growth phase, and a shrinkage phase. Therefore, it can be used to describe tumors having these three important phases. In Subsection \ref{subsec:parammodels} we discuss the role of the parameters in this general model and their interpretation.

\section{Origin and biophysical interpretation of our model} \label{sec:biophysicsexp}

\subsection{Parameters of the model}
\label{subsec:parammodels}

In this section we want to analyze the role of each parameter of the model in Eq. (\ref{eq:polgenV}). In subsection \ref{subsec:particularmodels} we already gained an understanding of the role played by each of the main parameters in that model. However, in the present section, we want to explore the effect of each parameter in more detail. Thus, we are open to the possibility of connecting each parameter with a biophysical process in tumor evolution. Since our model contains a free function $\phi(t)$, we stick to the models discussed in subsection \ref{subsec:particularmodels}. \\

\begin{figure}[H]
\centering
\includegraphics[scale=0.5]{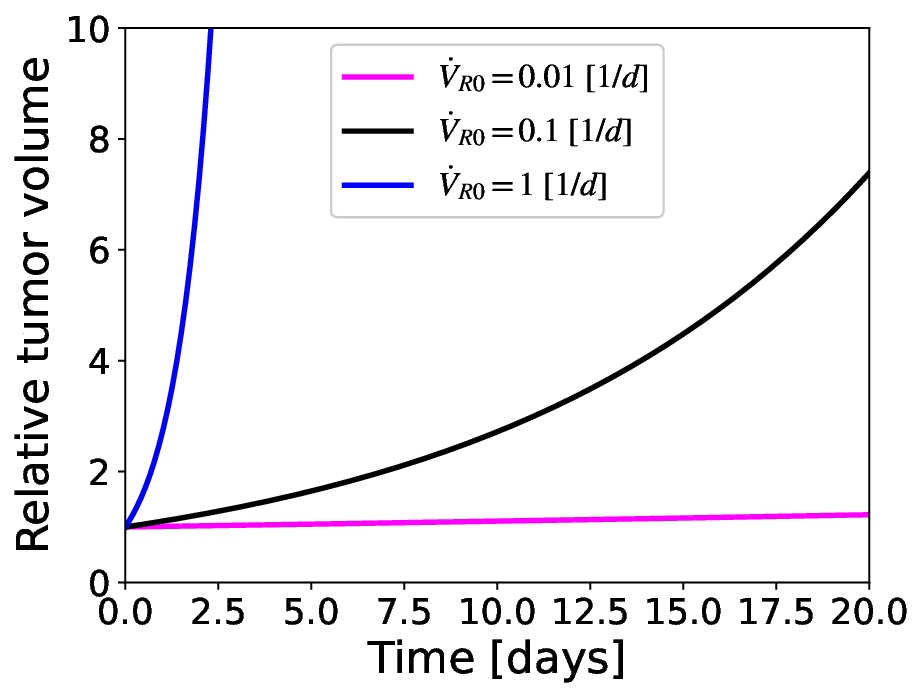}
\caption{Effect of $\dot{V}_{R0}$ our derived traditional exponential model $V_{R}(t)=e^{\dot{V}_{R0}t}$, which only contains the initial growth rate as a parameter}
\label{Fig:TradExp}
\end{figure}

\begin{figure}[H]
\centering
\includegraphics[scale=0.5]{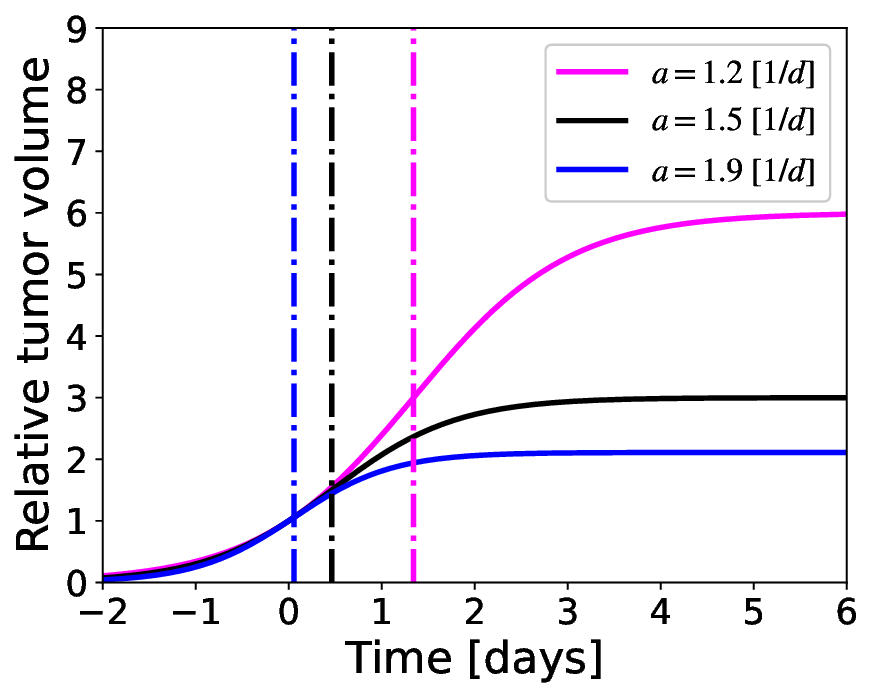}
\caption{Behavior of the Modified Logistic model of Eq. (\ref{eq:newlogistic}) for three different values of the parameter $a$ and $\dot{V}_{R0}=1\ [1/d]$. The dotted-dashed vertical lines represent the transition times for each case: $32.189$ hours for the magenta curve, $11.090$ hours for the black curve, and $1.330$ hours for the blue curve}
\label{Fig:ModiLogParam}
\end{figure}

We start with the quantity $\dot{V}_{R0}$, although is formally an initial condition. In our discussion of the particular models, we stated that the traditional exponential model is obtained when  $n=1$ and $a=\dot{V}_{R0}$. Therefore, the value of the initial growth rate has to be taken as a parameter to be determined from a fitting process. If we plot the exponential model $V_{R}(t)=e^{\dot{V}_{R0}t}$ using different values for $\dot{V}_{R0}$, we get Fig. \ref{Fig:TradExp}. As expected from the nature of this quantity, a high positive value produces fast growth at large times. Given that $\dot{V}_{R0}>0$, the initial growth factor tells us how fast the tumor will grow. It is also important to note that the tumor will always have an increasing behavior. This situation represents the case of an untreated tumor that is not even \emph{attacked} by the immune system. To sum up, $\dot{V}_{R0}$ determines how the tumor starts growing.
If we introduce the parameter $a$, as done in our Modified Logistic model, we get Eq. (\ref{eq:newlogistic}). In this model we have the two quantities: $\dot{V}_{R0}$ and $a$. We already know what $\dot{V}_{R0}$ does. Therefore, we analyze the effect of the parameter $a$. In Fig. \ref{Fig:ModiLogParam} we observe that $a$ flattens the curve and takes the relative volume to a constant value, as predicted by Eq. (\ref{eq:trantimemodilog}). In this figure we have also marked the \emph{transition time}. This time represents the moment when there is a transition from fast growth and to a stationary phase. This time is found by looking for the point at which $\ddot{V}_{R}=0$. Fig. \ref{Fig:ModiLogParam} also shows us that a large $a$ leads to a faster transition to the stationary phase and a smaller asymptotic relative volume. The transition time from growth to the stationary phase, as well as the value of the asymptotic volume, depends on the relation between $\dot{V}_{R0}$ and $a$. The expressions for both quantities can be seen in Eq. (\ref{eq:trantimemodilog}) and (\ref{eq:maxvn1}). Thus, we conclude that $a$ contains the effects of the immune system and a specific treatment leading to a global cytostatic terminal phase. In other words, the parameter $a$ avoids uncontrolled growth. \\

\begin{figure}[H]
\centering
\includegraphics[scale=0.5]{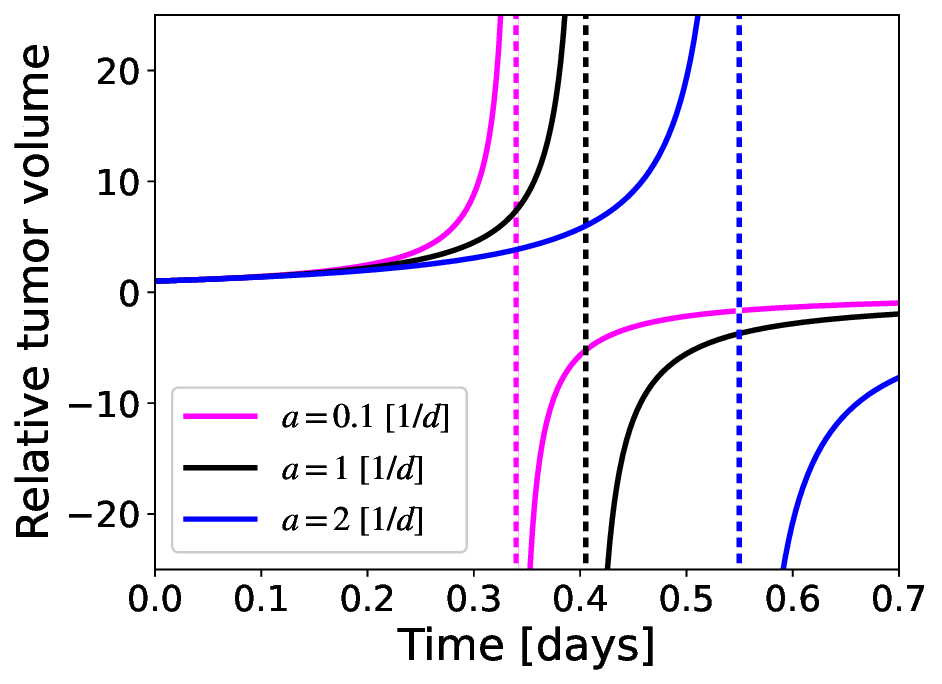}
\caption{Behavior of our Modified Logistic model of Eq. (\ref{eq:newlogistic}) when $\dot{V}_{R0}>a$ for three different values of $a$ and $\dot{V}_{R0}=3\ [1/d]$. The three vertical dashed lines represent the asymptotes of each function. In this case, the model is clearly valid from $t=0$ to the position of each asymptote}
\label{Fig:ModiLogasmall}
\end{figure}

Before moving to the other parameters, we want to show again that our Modified Logistic model can lead to growth characterized by the presence of a vertical asymptote. This asymptote makes the growth fast and corresponds to the case in which $\dot{V}_{R0}>a$. In other words, this situation happens when the value of the parameter $a$ containing information about cytostatic effects is not enough to counteract the initial growth driven by $\dot{V}_{R0}$. In Fig. \ref{Fig:ModiLogasmall} we see three curves showing tumor evolution when $\dot{V}_{R0}>a$.

\begin{figure}[H]
\centering
\includegraphics[scale=0.5]{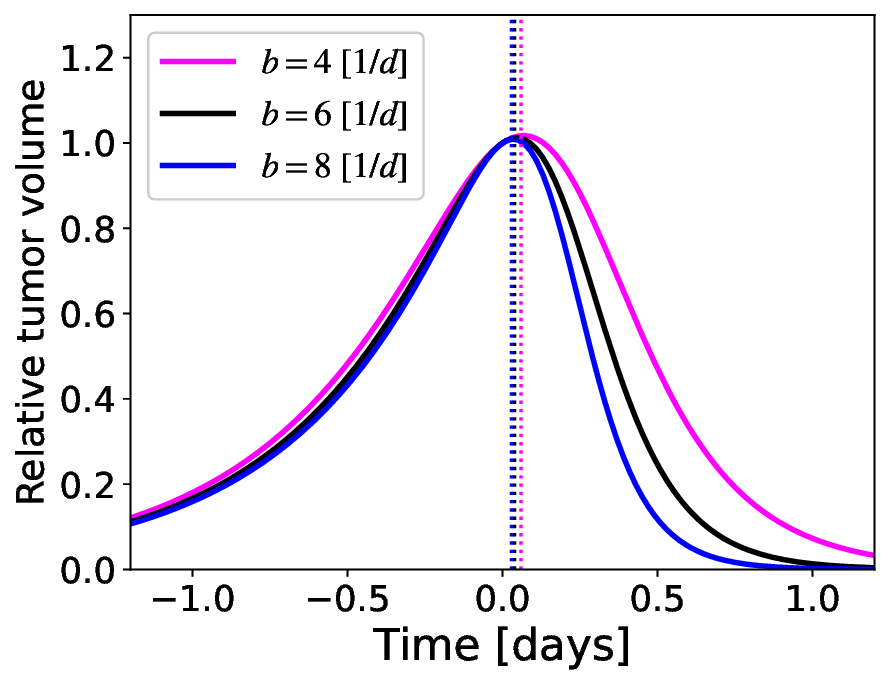}
\caption{Behavior of the model of Eq. (\ref{eq:trantimemodilog}) for three different values of the parameter $b$, $\dot{V}_{R0}=0.5\ [1/d]$ and $a=2\ [1/d]$. The dotted lines represent the position of the maximum for each function: $1.427$ hours for the magenta curve, $55.828$ minutes for the black curve, and $41.426$ minutes for the blue curve}
\label{Fig:AsymShrb}
\end{figure}

Now we talk about the effect of the parameter $b$. The model that allows us to easily understand the effect of $b$ is the one given in Eq. (\ref{eq:shrinkageab}) and called \emph{model with terminal asymmetric shrinkage}. As understood from this name, this model will show a late-times decreasing evolution. The reason for this decreasing phase can be understood from the form of Eq. (\ref{eq:shrinkageab}) since there is an exponential function in the denominator that grows with time. In Fig. \ref{Fig:AsymShrb} we observe the behavior of the model for three values of the parameter $b$. A large $b$ leads to an evolution that approaches zero in a shorter time. Thus, we can say that $b$ has a cytotoxic nature. This behavior was already clear in Subsection \ref{subsec:particularmodels}, where we separated our model into two groups: those with terminal shrinkage and those without terminal shrinkage. All models with terminal shrinkage are characterized by a non-zero $b$. We want to stress that the actual decreasing behavior depends on all the parameters since the exponential function $e^{bt}$ is multiplied by a factor $B$. To sum up, we can argue undoubtedly that $b$ has a cytotoxic effect as long as it is positive.

\begin{figure}[H]
\centering
\includegraphics[scale=0.5]{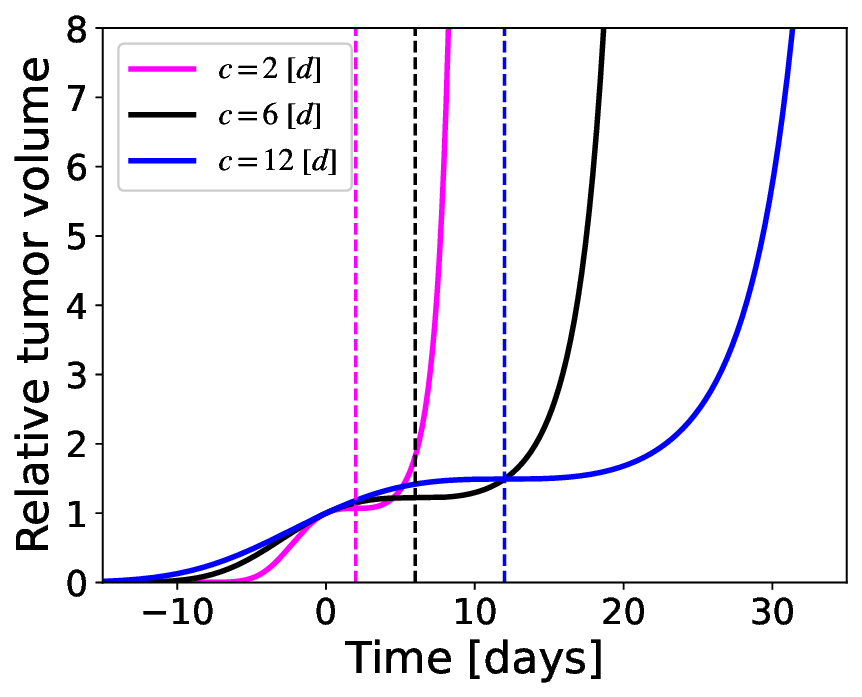}
\caption{Behavior of the model of Eq. (\ref{eq:genexpmodn}) for three different values of $c$ and $\dot{V}_{R0}=0.1\ [1/d]$, $n=3$. Each dashed vertical line represents the position of $t=c$. We observe that a large $c$ creates a longer near-stationary phase}
\label{Fig:GenExpParamc}
\end{figure}

Continuing with our analysis, we address the role of the parameter $c$. Looking at the general model of Eq. (\ref{eq:genexpmodn}) we see that it represents a shift of the function when $n>1$. However, it can induce more changes depending on the particular values of $a$, $b$, and $n$. For instance, in Fig. \ref{Fig:GenExpParamc} we see that $c>0$ shifts the function to the right, and that a larger $c$ leads to a longer stationary phase before the function grows rapidly to infinity. This result is due to the $c$ dependence of the Generalized Exponential model. In Eq. (\ref{eq:genexpmodn}) we can see that the function $(t-c)^{n}$ is multiplied by the factor that goes like $1/(-c)^{n-1}$. Then, a larger $c$ leads to a smaller factor in the exponent. \\

\noindent The generalized model with terminal shrinkage of Eq. (\ref{eq:genshr}) also allows us to see the effect $c$. We get a similar behavior as before. First, there is a horizontal shift depending on the value of $c$. Additionally, the near-stationary phase is modified by the chosen value for $c$.\\

\noindent We also discuss the effect of $c$ in the Generalized Modified Logistic model of Eq. (\ref{eq:genmodlog}). An interesting aspect of this model relies on the existence of two \emph{plateaux} in the evolution of $V_{R}(t)$. We see in Fig. \ref{Fig:GenModiLogParamc} how the properties of those plateaux change with the value of $c$. Namely, the clear separation and existence of the two plateaux are imperceptible as $c$ grows.

\begin{figure}[H]
\centering
\includegraphics[scale=0.5]{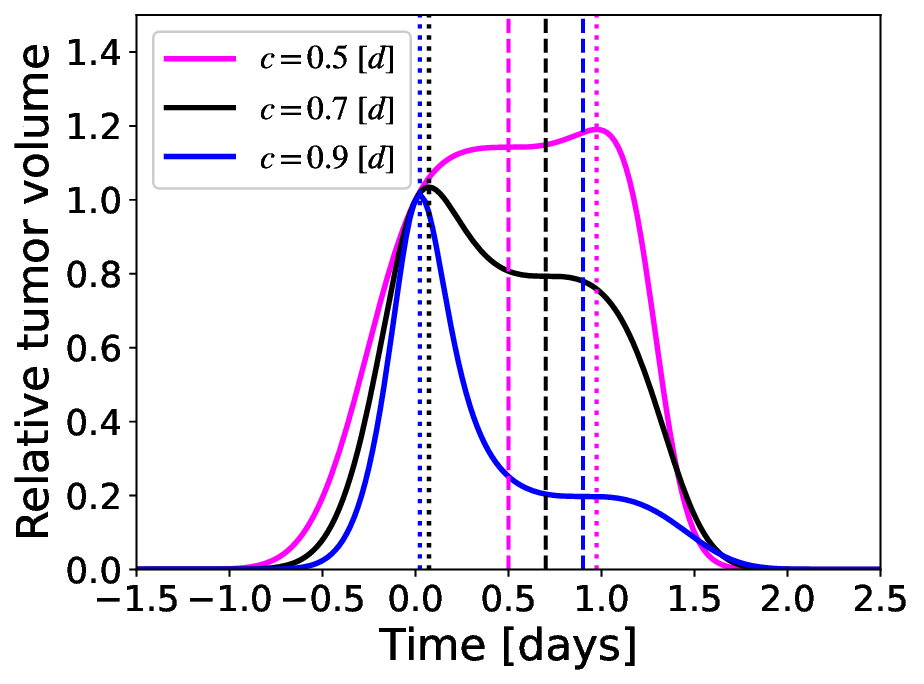}
\caption{Behavior of the model of Eq. (\ref{eq:genshr}) for three different values of $c$ and $\dot{V}_{R0}=1\ [1/d]$, $a=2\ [1/d^{3}]$, $b=4\ [1/d^{3}]$, $n=3$. Each dashed vertical line represents the position of $t=c$, and each dotted line represents the position of the maximum of each function. We clearly see that the value of $c$ modifies the evolution of the function and the properties of the region between growth and shrinkage}
\label{Fig:GenShrParamc}
\end{figure}

\begin{figure}[H]
\centering
\includegraphics[scale=0.5]{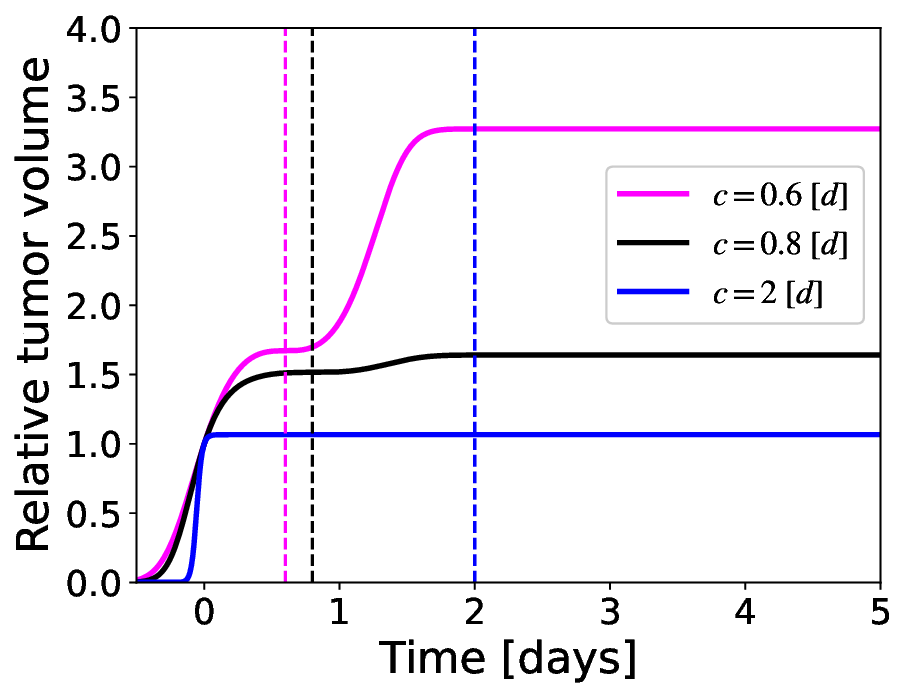}
\caption{Behavior of the model of Eq. (\ref{eq:genmodlog}) for three different values of $c$ and $\dot{V}_{R0}=3\ [1/d]$, $a=4\ [1/d^{3}]$ and $n=3$. Each dashed vertical line represents the position of $t=c$. We clearly see that the value of $c$ modifies the evolution of the function and the properties of the stationary regions}
\label{Fig:GenModiLogParamc}
\end{figure}

\begin{figure}[H]
\centering
\includegraphics[scale=0.5]{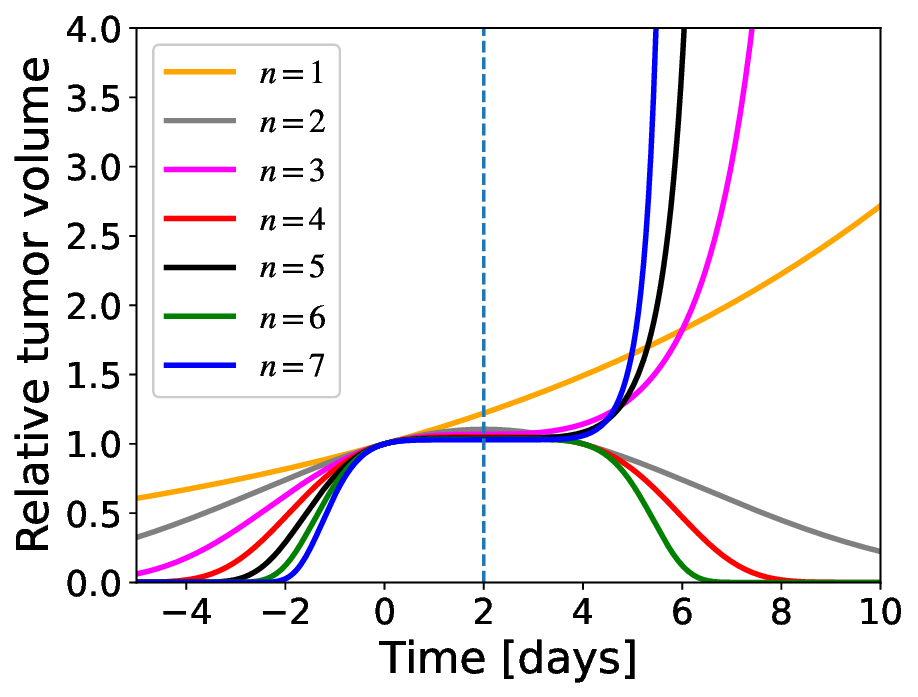}
\caption{Behavior of the model of Eq. (\ref{eq:genexpmodn}) for seven different values of $n$ and $\dot{V}_{R0}=0.1\ [1/d]$, $c = 2\ [d]$. The dashed vertical line represents the point $t=c=2\ [d]$. A parameter $n>0$ creates a near-stationary region. The evenness of the parameter $n$ also affects the evolution, as can be seen for the cases $n=2,4,6$. However, these last cases are inconsistent with our initial assumption of a positive parameter $a$ when $\dot{V}_{R0}>0$}
\label{Fig:GenExpParamn}
\end{figure}

\begin{figure}[H]
\centering
\includegraphics[scale=0.5]{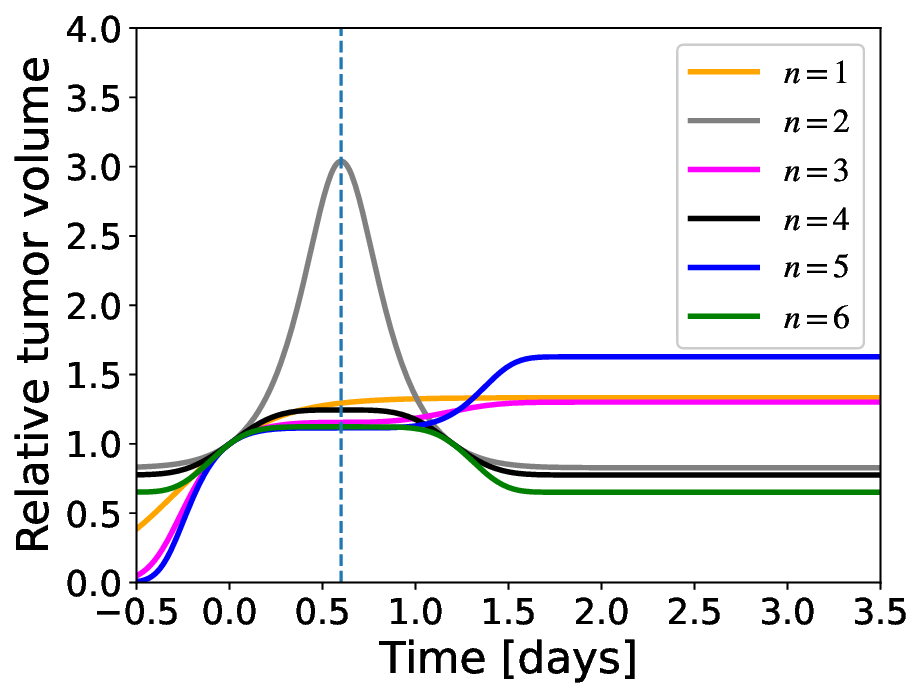}
\caption{Behavior of the model of Eq. (\ref{eq:genmodlog}) for six different values of $n$ and $\dot{V}_{R0}=1\ [1/d]$, $a=4\ [1/d^{n}]$ and $c = 0.6\ [d]$. The dashed vertical line represents the point $t=c=0.6\ [d]$. A parameter $n>1$ creates a near-stationary region. The evenness of t$n$ also affects the evolution, as can be seen for the cases $n=2,4,6$}
\label{Fig:GenModiLogParamn}
\end{figure}

Now we discuss the effect of the parameter $n$, the only dimensionless parameter in the model of Eq. (\ref{eq:polgenV}). In Fig. \ref{Fig:GenExpParamn} we see that $n$ creates a near-stationary phase around the value of $c$ in the model (\ref{eq:genexpmodn}). For $n=1$ there is only growth. However, as long as $n>1$, there is a near-stationary phase. During this phase, the change in $V_{R}$ is imperceptible and hence very small. Subsequently, the function grows to infinity or goes to zero depending on the value of $n$. We see that a large $n$ indicates the sharpest growth at late times when the parameter $n$ is odd. However, the evenness of $n$ is also important. If $n$ is even, the function decreases at large times. If $n$ is even in the model (\ref{eq:genexpmodn}), we get a negative value for the parameter $a$, which is inconsistent with our initial assumptions. We can observe something similar in Fig. \ref{Fig:GenModiLogParamn} for the model in Eq. (\ref{eq:genmodlog}). For this generalized modified logistic model the terminal behavior of $V_{R}$ and its intermediate near-stationary region are also affected by the evenness of the parameter $n$.\\

\begin{figure}[H]
\centering
\includegraphics[scale=0.5]{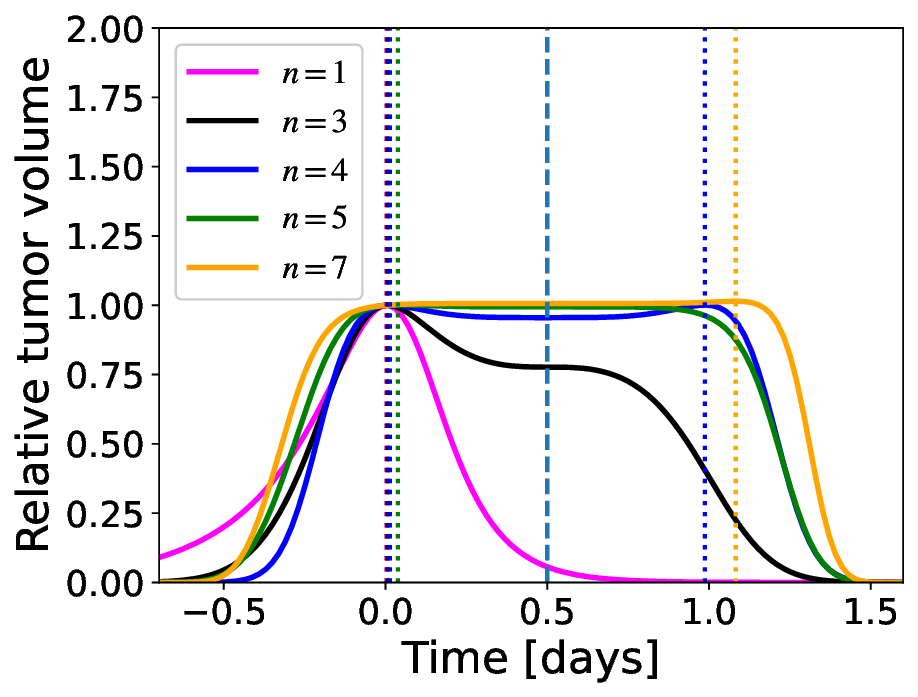}
\caption{Behavior of the model of Eq. (\ref{eq:genshr}) for five different values of $n$ and  $\dot{V}_{R0}=0.1\ [1/d]$, $a=4\ [1/d^{n}]$, $b=8\ [1/d^{n}]$. The dashed vertical line represents the point $c=2\ [d]$, and the dotted lines represent the position of the maxima for each function; the case $n=4$ displays two symmetric maxima around $t=c$. It is clear that a parameter $n>0$ creates a near-stationary region, and its duration increases with the value of $n$}
\label{Fig:GenShrParamn}
\end{figure}

On the other hand, in Fig. \ref{Fig:GenShrParamn} we see the impact of the parameter $n$ in the model described by Eq. (\ref{eq:genshr}). Once again, we see that there is a near-stationary region as long as $n>1$. Moreover, this phase tends to last longer when $n$ increases. Regarding the shrinkage phase, the decay is sharper when $n$ is large. Therefore, we can say that the parameter $n$ might both \emph{cytostatic} and \emph{cytotoxic} effects on tumor evolution.  

\subsection{Second-order nature of the model}

In this subsection we discuss the second-order nature of our model in more detail. As we showed before, most models for tumor evolution come from first-order differential equations. This fact makes the interpretation of the differential equation clear in terms of rates. Moreover, those models require only the knowledge of the volume at an initial time $t_{0}$. However, we know that dynamical models often come from second-order differential equations. Since the evolution of tumors is in-depth related to the motion of molecules and dynamical processes, we expect the resulting volume to be described by a function that is derived from a second-order differential equation. Being the volume a macroscopic descriptor, it is reasonable to think that it can be derived from the microscopic dynamics. Therefore, a natural form for the differential equation governing $V$ would have the form
\begin{equation}
    \frac{d^{2}V}{dt^{2}}=F\left(V,\frac{dV}{dt},t\right),
    \label{eq:dynamicaleq}
\end{equation}
where $F$ is any function of the variables shown in the equation. This form is precisely the one we introduced in Eq. (\ref{eq:mostgendeV}).\\

On the other hand, it is also reasonable to expect that the evolution of the tumor volume depends on its initial rate of change. Any dynamical problem requires the knowledge of the initial rate of change of its variables. Hence, our model is compatible with this physically-motivated argument. Moreover, the necessity of knowing $\dot{V}(t_{0})$ urges us to design more accurate experiments that allow us to get information about this initial rate of change. Thus, not only our model is novel but also lights the way for new experiments in the area of cancer research. In the following subsections we show how Eq. (\ref{eq:mostgendeV}) can be written to understand its origin and give interpretations to our findings in the previous sections.

\subsection{Sturm-Liouville problem}

For a given variable $y$ depending on $x$, the Sturm'Liouville differential equation is given by
\begin{equation}
    {\cal L}y = \frac{d}{dx}\left[p(x) w(x)\frac{dy}{dx}\right]+q(x)y=-\lambda w(x)y,
    \label{eq:SLeq}
\end{equation}
where $p(x)$ and $q(x)$ are functions of the variable $x$, over a region of interest $x\in [x_{i},x_{f}]$. Here the constant $\lambda$ corresponds to the eigenvalue and $w(x) > 0$ is known as the weight function. Additionally, the operator ${\cal L} y$ is called a self-adjoint operator \citep{Arfken}.\\

In order to write our Eq. (\ref{eq:mostgendeV}) as a Sturm-Liouville problem, we start from Eq. (\ref{eq:Vandphi})
\begin{equation}
\frac{d^{2}V}{d\phi^{2}}-\frac{2}{V}\left(\frac{dV}{d\phi}\right)^{2}+(a-b)\frac{dV}{d\phi}+abV=0,
\end{equation}
which was obtained by changing the independent variable from $t$ to $\phi$. Then, we introduce the variable $y=\frac{1}{V}$. Thus, we arrive at the differential equation
\begin{equation}
    \frac{d}{d\phi}\left[\frac{dy}{d\phi}+(a-b)y\right]=aby.
\end{equation}
It is straightforward to show that this equation can be written in the following form
\begin{equation}
    \frac{d}{d\phi}\left[e^{(a-b)\phi}\frac{dy}{d\phi}\right]=abe^{(a-b)\phi}y,
    \label{eq:SLTeq}
\end{equation}
where the constant $\lambda = -a b$ corresponds to the eigenvalue.

\subsection{Continuity equation}

Besides having the property of being written as a Sturm-Liouville differential equation, our model (\ref{eq:mostgendeV}) can also be written as a continuity equation. This form is particularly interesting it opens up the possibility of understanding our model in terms of fluid dynamics. If we start from (\ref{eq:SLTeq}) and go back to the variable $t$, we get the form
\begin{equation}
    \frac{d}{dt}\left[\frac{dV}{d\phi}\frac{e^{(a-b)\phi}}{V^{2}\dot{\phi}}\right]+\frac{ab\dot{\phi}e^{(a-b)\phi}}{V}=0.
    \label{eq:compmostgendeV}
\end{equation}
Since the expression inside the derivative of the first term is a function of time, we can define
\begin{equation}
    Q(t)=\frac{dV}{d\phi}\frac{e^{(a-b)\phi}}{V^{2}\dot{\phi}}.
    \label{eq:defQ}
\end{equation}
Moreover, we can also introduce the function
\begin{equation}
    \Phi=\frac{ab\dot{\phi}e^{(a-b)\phi}}{V}.
    \label{eq:defPhi}
\end{equation}
Thus, it is clear that the differential equation (\ref{eq:compmostgendeV}) can be written as
\begin{equation}
    \frac{\partial Q}{\partial t}+\Phi=0.
\end{equation}
This form resembles a continuity equation where $Q$ and $\Phi$ come from integrals of the form
\begin{equation}
Q=\int_{V}\rho\ dV, \ \ \Phi=\int_{S}\mathbf{J}\cdot\mathbf{dS}.
\end{equation}
Here, the quantity $\rho$ represents the volume density of a given biophysical concept, and $\mathbf{J}$ represents the associated current density.
As a consequence, our model can be interpreted in terms of a conservation law. In the context of cancer, we have the dynamics of fluids coming in and out of the tumor. Hence, the existence of a continuity equation is unsurprising. It is also interesting to note that the \emph{flux} $\Phi$ is multiplied by the parameter $b$. As we saw in the previous sections, the parameter $b$ is related to terminal-shrinkage effects as long as $\phi(t)$ is positive at large times. Hence, if $b=0$ there is no outward flux and the volume does not decrease at large times. In that case, we either have uncontrolled growth or a maximum volume $V^{\infty}$ at large times. Additionally, the case $b=0$ shows us that $Q$ is a conserved quantity or a constant of motion since
\begin{equation}
    \frac{\partial Q}{\partial t}=0.
\end{equation}
Therefore, we say that this quantity is a constant $Q(t)=C$.
In fact, for the case of a vanishing $b$, Eq. (\ref{eq:compmostgendeV}) becomes
\begin{equation}
    \frac{d}{dt}\left[\frac{dV}{d\phi}\frac{e^{a\phi}}{V^{2}\dot{\phi}}\right]=0\ \longrightarrow \frac{dV}{d\phi}\frac{e^{a\phi}}{V^{2}\dot{\phi}}=C,
\end{equation}
which can be easily solved and has the solution
\begin{equation}
    V(t)=\frac{1}{D+(C/a)e^{-a\phi(t)}},
    \label{eq:Vb0}
\end{equation}
where $D$ is an integration constant. This Eq. (\ref{eq:Vb0}) leads all the exponential and modified logistic models discussed in the previous sections. We conclude then that all the models without terminal shrinkage come from a conserved quantity $C$. This interesting result will be analyzed in future work where we will study the spatiotemporal microscopic dynamics of tumor evolution. We are certain that this conclusion opens up questions that must be studied theoretically and experimentally by more research groups within the areas of Biology, Mathematics, Physics and Chemistry.\\

We want to close this section by discussing the opposite case, the one in which $a=0$. In such a case, there is also a conserved quantity
\begin{equation}
    \frac{d}{dt}\left[\frac{dV}{d\phi}\frac{e^{-b\phi}}{V^{2}\dot{\phi}}\right]=0\ \longrightarrow \frac{dV}{d\phi}\frac{e^{-b\phi}}{V^{2}\dot{\phi}}=\tilde{C}.
\end{equation}
With this form, we find the expression for the volume
\begin{equation}
    V(t)=\frac{1}{\tilde{D}-(\tilde{C}/b)e^{b\phi(t)}},
    \label{eq:Va0}
\end{equation}
where $\tilde{D}$ is an integration constant. This function has the property of being decreasing for large times. Therefore, it corresponds to a terminal-shrinkage model as long as the pole (vertical asymptote) in the function lies at `negative times' (before the initial time $t_{0}$). This result supports our findings regarding the nature of the parameter $b$: it promotes the shrinkage of $V$.\\


\section{Methods}\label{sec:methods}

In this \emph{Methods} section we want to explain how we obtained the parameters from the experimental data sets used throughout the paper. As stated in Appendix \ref{secA2}, we used the \verb|curve_fit| tool in Python. Since the cases discussed in subsec. \ref{subsec:particularmodels} present all characteristic features, we paired data sets with models having similar properties from a graphical point of view. Such properties are, for instance, terminal shrinkage or the existence of plateaux. In our analysis we used reasonable initial-guess values for the parameters given the discussion in subsection \ref{subsec:parammodels} and the trend observed in each data set. Additionally, we used a large \verb|maxfev| number of the order of thousands to have a large number of iterations. For each of the fitted data, we made sure that $R^{2}\gtrsim0.7$ and the standard deviations of the parameters were small (a relative error smaller than around $0.5$). These requirements led us to favor some models against others. For instance, in the BEV/DDP/PTXHigh group, we used the simple exponential model instead of our symmetric shrinkage model for two of the data sets. The reader is invited to check the fits shown in this paper and test our model using other experimental studies as well.

\section{Discussion and Conclusions}\label{sec:Con}

We introduced and analyzed a new model that can be used to describe the evolution of biological populations. In this paper we showed an application to tumor evolution. The model that we introduced is innovative in several aspects. First, we hypothesize that a second-order model can be used to account for different stages in tumor evolution: growth, quasi-stationary phases, and shrinkage. Second, the parameters in our model have a straightforward interpretation and can be used to link medical/therapeutic concepts to the mathematical description (see subsec. \ref{subsec:parammodels}, where we study the effect of each parameter in detail). The parameters of our model combine to produce specific patterns; however, we can undoubtedly say that the number $c$ marks the point of transition between two regions with different concavity, $n$ can create longer regions of quasi-stationary behavior, and $b$ determines the existence of terminal shrinkage. Third, we explore interesting properties of our mathematical formulation that exhibit a possible origin of our model in terms of a continuity equation. Fourth, we show that some classical models can be derived from our general formulation. Moreover, we explicitly show that our model(s) can be used to fit experimental data sets with diverse patterns. Our model does not restrict to growth or stationary regions as found in the traditional models; instead, we can account for a variety of phases and transitions that are present in tumor evolution. In Appendix \ref{secA2} we included the plots of all the 85 fits of the experimental data sets used to test our general model. We find particularly interesting those sub-models presenting growth and shrinkage. Since most treatments attempt to reduce the volume of a tumor, we find it beneficial to have a model describing such a phenomenon. As discussed in subsection \ref{subsec:particularmodels}, terminal shrinkage is present when the parameter $b$ is different from zero. Thus, we can connect the effectiveness of a given treatment with $b$ and other parameters of our model. Another important aspect that we want to highlight is the interpretation of the function $\phi(t)$. We showed that its form drastically determines the evolution of $V_{R}$. So, we argue that this function $\phi(t)$ is connected to the response of the biological system to some specific treatment and external factors, and we might see it as a \emph{biological clock} (a similar idea in different contexts was discussed in \cite{Gillooly} and \cite{Castorina}).  \\

Even though our model proves to be successful in the description of tumor evolution, further work is needed to test our formulation with more experimental data sets. It is necessary to show that our model can describe experiments performed with different systems and under different conditions. Moreover, a detailed study using experiments, data analysis, and theoretical considerations is needed to establish a clear link among the parameters of our model and medical concepts involved in cancer treatments. In particular, it would be extremely impactful to check the predictivity of our formulation. That is, given a few initial experimental data points, it would be remarkable to predict the late-times state of a tumor. Thus, physicians would be able to assess the effectiveness of a specific treatment. Relatedly, future experimental studies should also include a detailed analysis of the uncertainties present in the measurements. These uncertainties would restrict the spectrum of models that are suitable for fitting experimental data and a more accurate description would be possible.\\

Another aspect that we should consider is the deterministic nature of our formulation. We know that there are stochastic processes in these biological systems. Therefore, small fluctuations should also be considered in future studies. We claim that our formulation can be used as a background model in the study of tumors, cell colonies, and other types of biological populations. Work in this direction might be fruitful in the understanding of cancer. Additionally, another interesting study that can be carried out entails the possible values of the parameter $n$. In our study, we focused on integer values; however, there are no restrictions on this parameter as long as it is not of the form $\frac{p}{2q}$, with $p$ any odd number and $p\in\mathbb{N}$.

\backmatter

\bmhead{Acknowledgments}

We would like to thank Professor Paolo Ubezio for sharing the experimental data sets from the references \cite{Ubezio} and \cite{falcetta2017modeling}. F.D.L-C is supported by the Vicerrectoría de Investigación y Extensión - Universidad Industrial de Santander, under Grant No. 3703.

\section*{Declarations}

{\bf Conflict of interest:} The authors declare that they have no conflict of interest.

\appendix
\section{Polynomial time dependence}\label{secA1}

In this Appendix we discuss a further case for the choice of $\phi(t)$. As we exemplified above, the form of $\phi(t)$ is a key factor in the evolution of the tumor volume. So far we have focused on the case of a power-law dependence $\phi(t)=(t-c)^{n}$. This choice proved to be effective in the description of the majority of data sets found in
 \cite{falcetta2017modeling} and \cite{Ubezio}. However, there is no restriction on the form of $\phi(t)$ and we are free to choose a more complex form. In particular, we can take a polynomial in $t$. The choice of a polynomial function allows us to describe small variations of $V$ during its global evolution. \\

 \noindent Given the number of free parameters that might be involved in the case of a polynomial $\phi(t)$, we decided to discuss such a case in this appendix. Instead, we focused on models with a small number of free parameters with a more straightforward interpretation. Although the polynomial case is not the main result of our work, we found it important to present this situation in this article. \\ 

\noindent {\bf \emph{ Polynomial exponential model:}} \\
 
 In our analysis of the experimental results in \cite{falcetta2017modeling} and \cite{Ubezio} we found that the sets fitted with the generalized exponential model were also fitted with an exponential $\phi(t)$. That is, we might have a \emph{polynomial exponential model}. However, the fit looked quite fine-tuned and some of the standard deviations of the fitted parameters were extremely large. Therefore, we took the simplest models in our fitting process for the majority of experimental data sets. However, three data sets show particular evolution.  These three data sets were fitted with the case that we called
\emph{polynomial exponential model}. This model can be derived from the main equation (\ref{eq:mostgenV}) and it can be written as
\begin{equation}
    V_{R}(t)=e^{\phi_{pol_{n}}(t)},
    \label{eq:polexpmod}
\end{equation}
where
\begin{equation}
   \phi_{pol_{n}}(t)=\alpha t+\alpha_{2} t^{2}+\alpha_{3}t^{3}
+...+\alpha_{n}t^{n},
\label{eq:polphi}
\end{equation}
and the parameter $a$ has been absorbed in $\phi_{pol_{n}}(t)$\\

\noindent {\bf \emph{Interpretation of the polynomial exponential model:}} \\

\begin{figure}[H]
\centering
\includegraphics[scale=0.6]{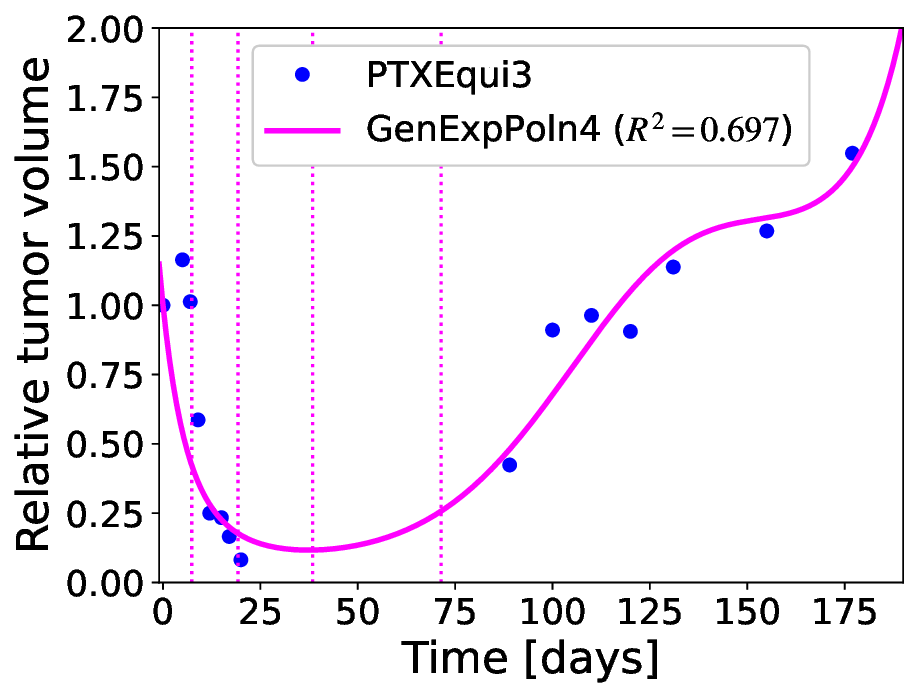}
\caption{Result of the fitting process: the dots represent the experimental data \emph{PTXEqui3} from \cite{falcetta2017modeling} and the continuous line represents the best fit using (\ref{eq:polexpmod}). The vertical lines mark the characteristic times $t_{i}$ defined in (\ref{eq:newpolphi})}
\label{Fig:PolyModel}
\end{figure}

Although the possibility of having a polynomial function $\phi(t)$ to describe the evolution of a tumor looks fine-tuned and perhaps unrealistic, we want to propose an interpretation that opens up new directions in the study of tumors. To avoid the presence of unexplained coefficients with very small values in the polynomial, we rewrite Eq. (\ref{eq:polphi}) as
\begin{equation}
\phi(t)=\beta_{1}\left(\frac{t}{t_{1}}\right)+\beta_{2}\left(\frac{t}{t_{2}}\right)^{2}+\cdots+\beta_{i}\left(\frac{t}{t_{i}}\right)^{i}+\cdots+\beta_{n}\left(\frac{t}{t_{n}}\right)^{n},
\label{eq:newpolphi}
\end{equation}
where $\beta_{i}$ can be either $+1$ or $-1$, and $t_{i}$ are characteristic times. The presence of the parameters $\beta_{i}$ removes the fine-tuning problem of adjusting very small numbers. Then, the relevance of any term is determined by the characteristic time $t_{i}$. If we have that $t_{n}>t_{n-1}>\cdots >t_{2}>t_{1}$, we can associate the relevance of each term in the polynomial to a specific time when a given effect starts taking place. Thus, we can link the evolution of $V$ to some medical events that are known or expected by physicians. In Fig. \ref{Fig:PolyModel} we show an example of a data set that was fitted with our polynomial exponential model. In this case, the degree of the polynomial is equal to $4$. In Appendix \ref{secA2} we present other two cases where we used the polynomial exponential model.

\section{Fitting all the data sets}\label{secA2}
In this Appendix we want to show all the fits of the experimental data sets found in references \cite{falcetta2017modeling}
and \cite{Ubezio}. Here we use the same acronyms found in those references, whose meaning is explained at the end of the Introduction. Our purpose with this appendix is twofold: on the one hand, we want to show the power of our proposal; on the other hand, we want to make clear the different phases that can be observed in tumor evolution. This last aspect is extremely relevant in the understanding of tumors and their response to several treatments. All fits shown throughout the paper were found using the \verb|curve_fit| option in \emph{Python}. To make the plots look more organized, we defined some labels that stand for each model described in this paper. We defined this notation as follows:\\
\begin{itemize}
    \item ModiLogi: Modified Logistic model of Eq. (\ref{eq:newlogistic}).
\item SimpExp: Traditional exponential model.
\item GenExpn: Generalized exponential model of degree $n$ of Eqs. (\ref{eq:genexpmodn}) or (\ref{eq:genexpmodnnoc}) depending on the value of $c$.
\item GenModiLogin: Generalized modified logistic model of degree $n$ of Eqs. (\ref{eq:genmodlog}) or (\ref{eq:genmodlognoc}) depending on the value of $c$.
\item SymShr: Symmetric shrinkage model of Eq. (\ref{eq:symmshr}).
\item AsymShr: Asymmetric shrinkage model of Eq. (\ref{eq:shrinkageab}).
\item GenShr: Generalized shrinkage model of degree $n$ of Eqs. (\ref{eq:genshr}) or (\ref{eq:genshrnoc}) depending on the value of $c$.
\item GenExpPoln: Generalized exponential polynomial model of degre $n$ of Eq. (\ref{eq:polexpmod}).
\end{itemize}
It is important to stress that the word \emph{degree} stands for the power used in $\phi(t)=(t-c)^{n}$. The only case in which a full polynomial of degree $n$ is used corresponds to the GenExpPol model.\\

The following figures show the versatility of our formulation and its usefulness in describing different phases in tumor evolution. There are two findings that we want to highlight. First, all the \emph{control} sets are fitted with our modified logistic model of Eq. (\ref{eq:newlogistic}). Second, data sets within the same \emph{family} do not necessarily have identical evolution. As expected, the response to a specific treatment depends on the living system under study; in medical terms, the effectiveness of drug treatment will depend on the patient. Thus, our formulation opens up the possibility of connecting some parameters of our model to the human body response in terms of biological processes.

\begin{figure}[H]
\centering
\includegraphics[scale=0.41]{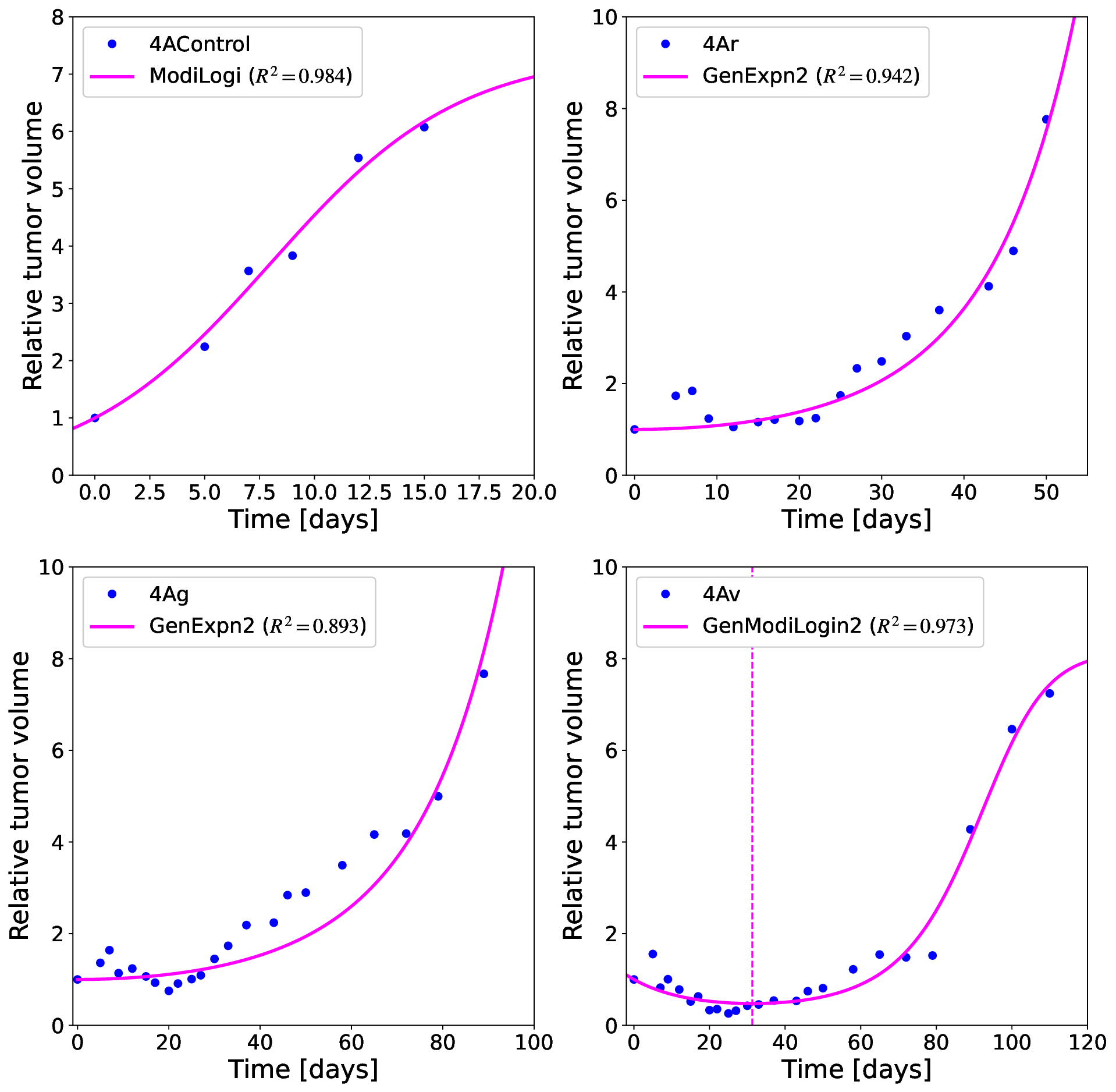}
\caption{Fits of the data sets named 4A in \cite{Ubezio}. The points represent the experimental values and the continuous lines represent the best fits. The dashed vertical lines in the last panel (4Av) marks the position of $c$}
\label{Fig:All4A}
\end{figure} 

\begin{figure}[H]
\centering
\includegraphics[scale=0.41]{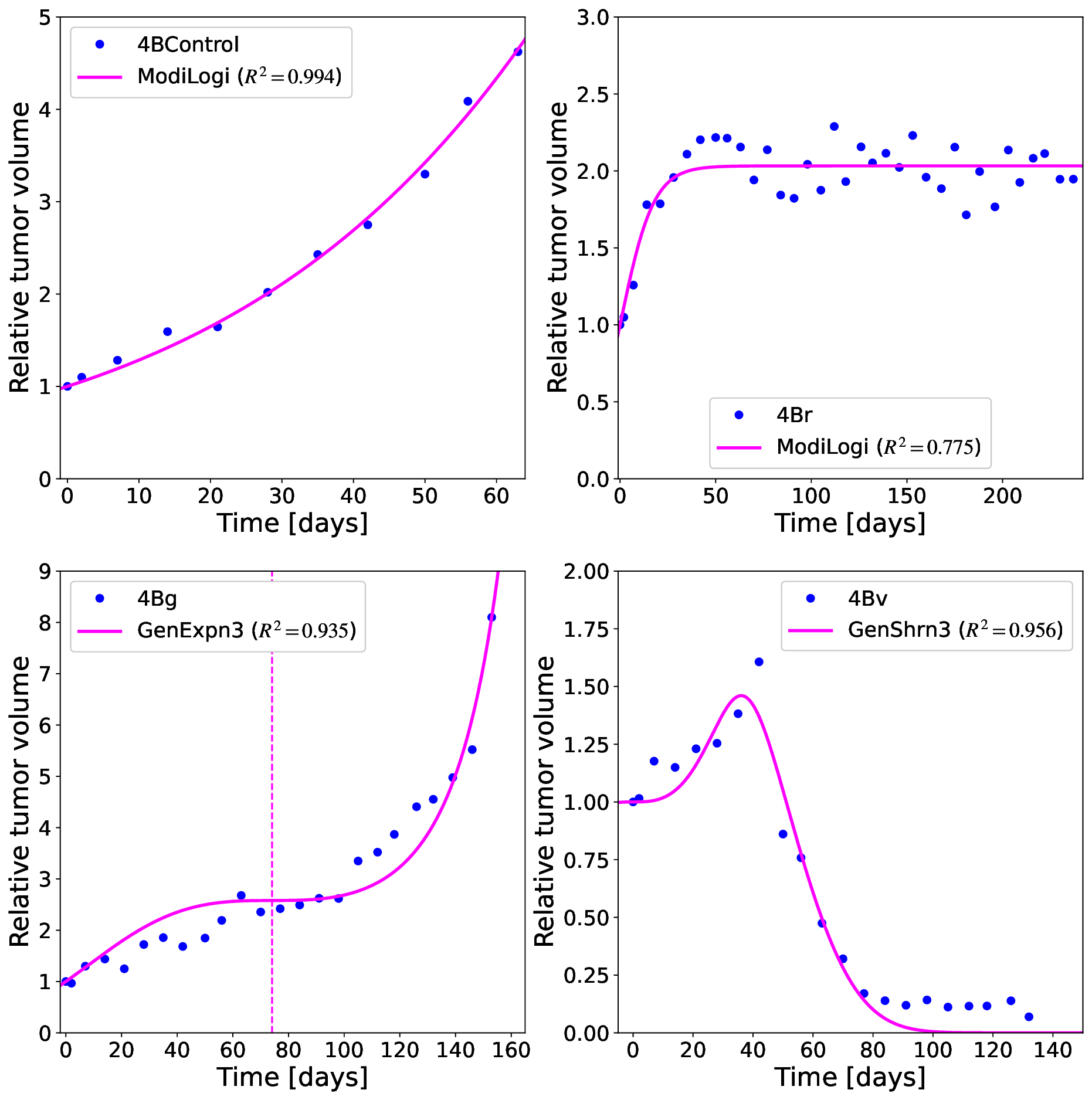}
\caption{Fits of the data sets named 4B in \cite{Ubezio}. The points represent the experimental values and the continuous lines represent the best fits. The dashed vertical line in 4Bg marks the position of $c$}
\label{Fig:All4B}
\end{figure} 

\begin{figure}[H]
\centering
\includegraphics[scale=0.39]{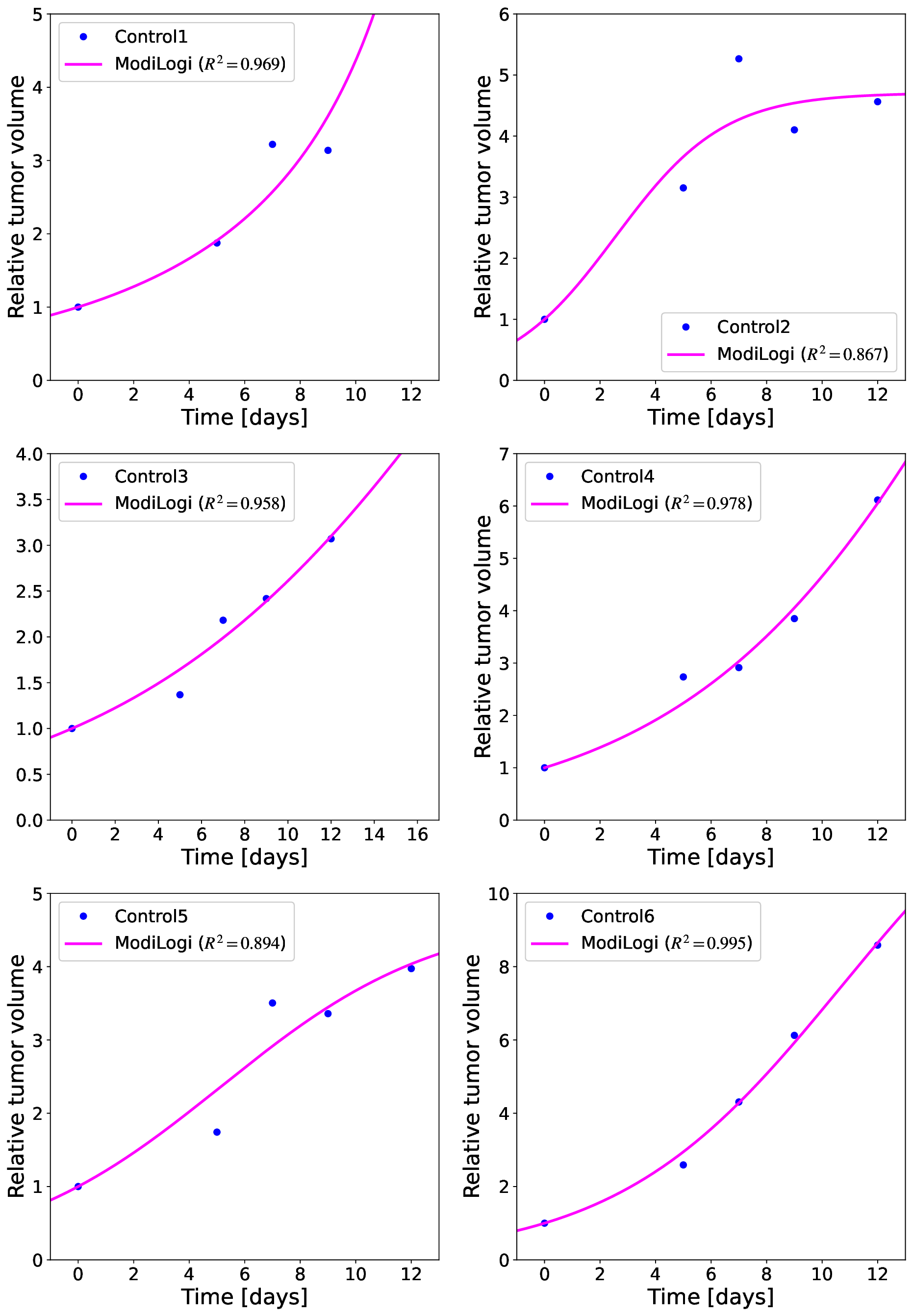}
\caption{Fits of the data sets named Control in \cite{falcetta2017modeling}. The points represent the experimental values and the continuous lines represent the best fits. All these data points are fitted using our modified logistic model of Eq. (\ref{eq:newlogistic})}
\label{Fig:AllControl}
\end{figure} 

\begin{figure}[H]
\centering
\includegraphics[scale=0.4]{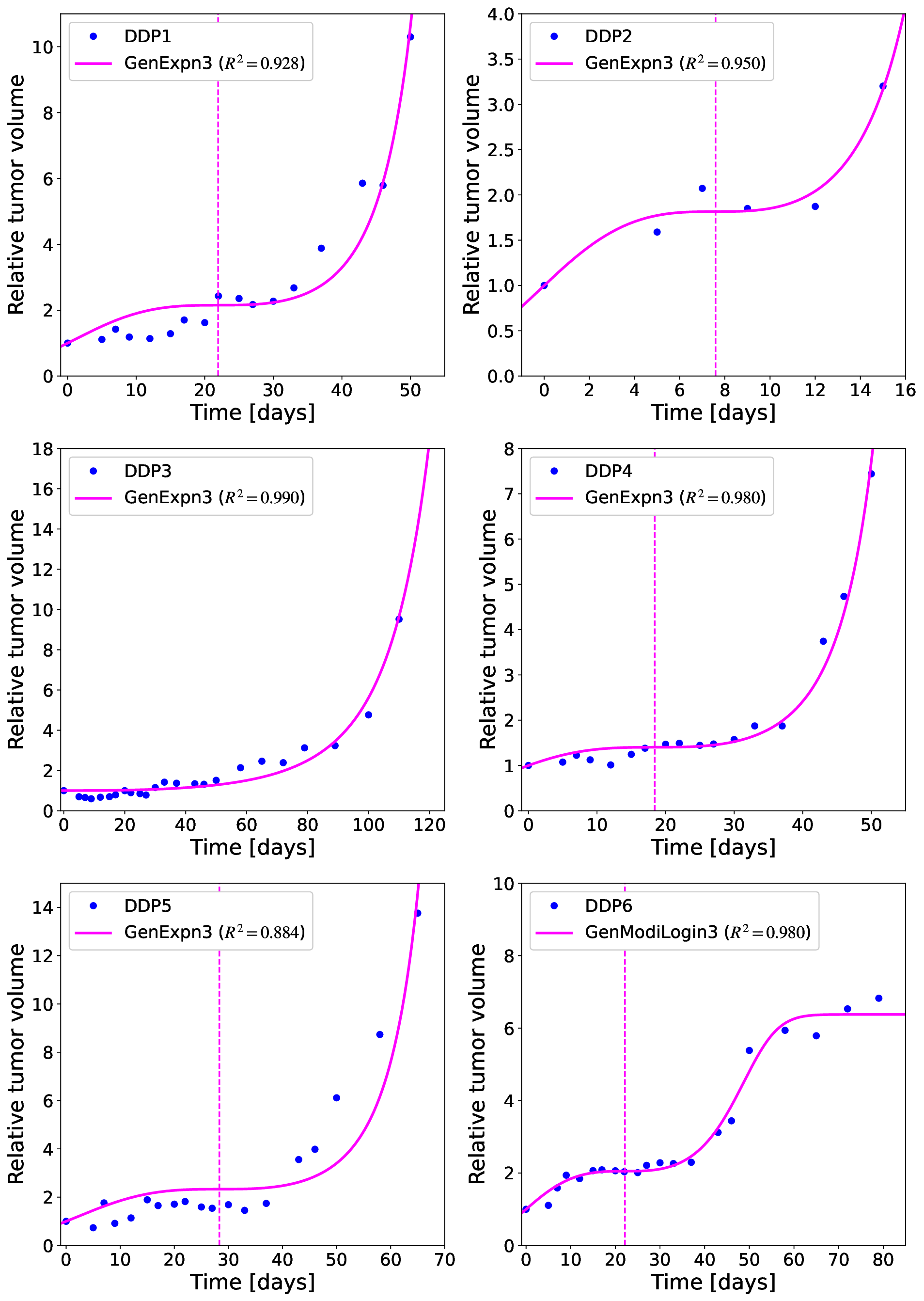}
\caption{Fits of the data sets named DDP in \cite{falcetta2017modeling}. The points represent the experimental values and the continuous lines represent the best fits. The dashed vertical lines in some plots mark the position of $c$. We observe that none of the models predict a vanishing volume in the far future}
\label{Fig:AllDDP}
\end{figure} 

\begin{figure}[H]
\centering
\includegraphics[scale=0.4]{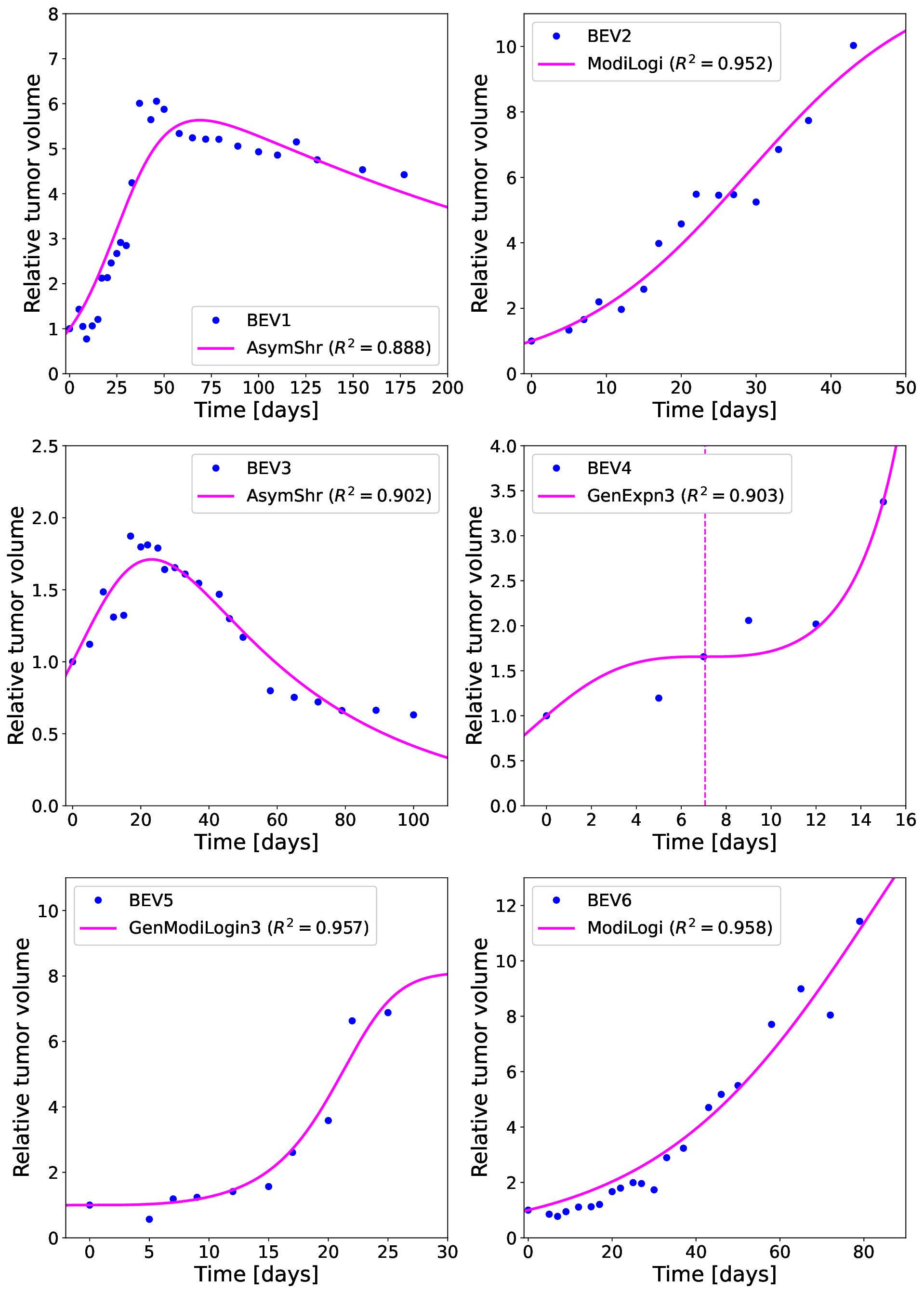}
\caption{Fits of the data sets named BEV in \cite{falcetta2017modeling}. The points represent the experimental values and the continuous lines represent the best fits. The dashed vertical line in BEV4 marks the position of $c$. We observe that only in BEV1 and BEV3 there is volume shrinkage when $t\rightarrow\infty$}
\label{Fig:AllBEV}
\end{figure} 

\begin{figure}[H]
\centering
\includegraphics[scale=0.4]{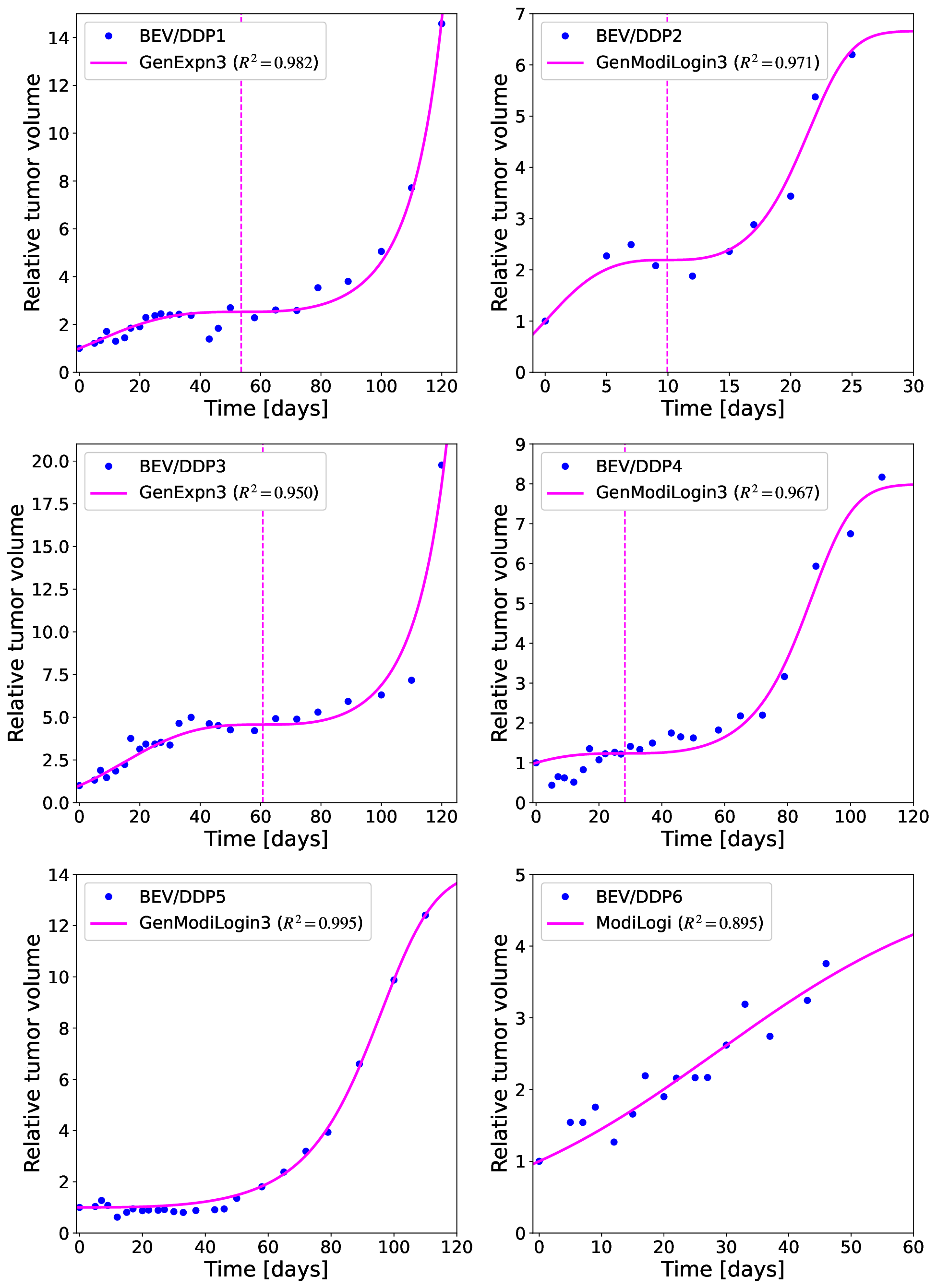}
\caption{Fits of the data sets named BEV/DDP in \cite{falcetta2017modeling}. The points represent the experimental values and the continuous lines represent the best fits. The dashed vertical lines in some plots mark the position of $c$. We observe that none of the models predict a vanishing volume in the far future}
\label{Fig:AllBEVDDP}
\end{figure} 

\begin{figure}[H]
\centering
\includegraphics[scale=0.4]{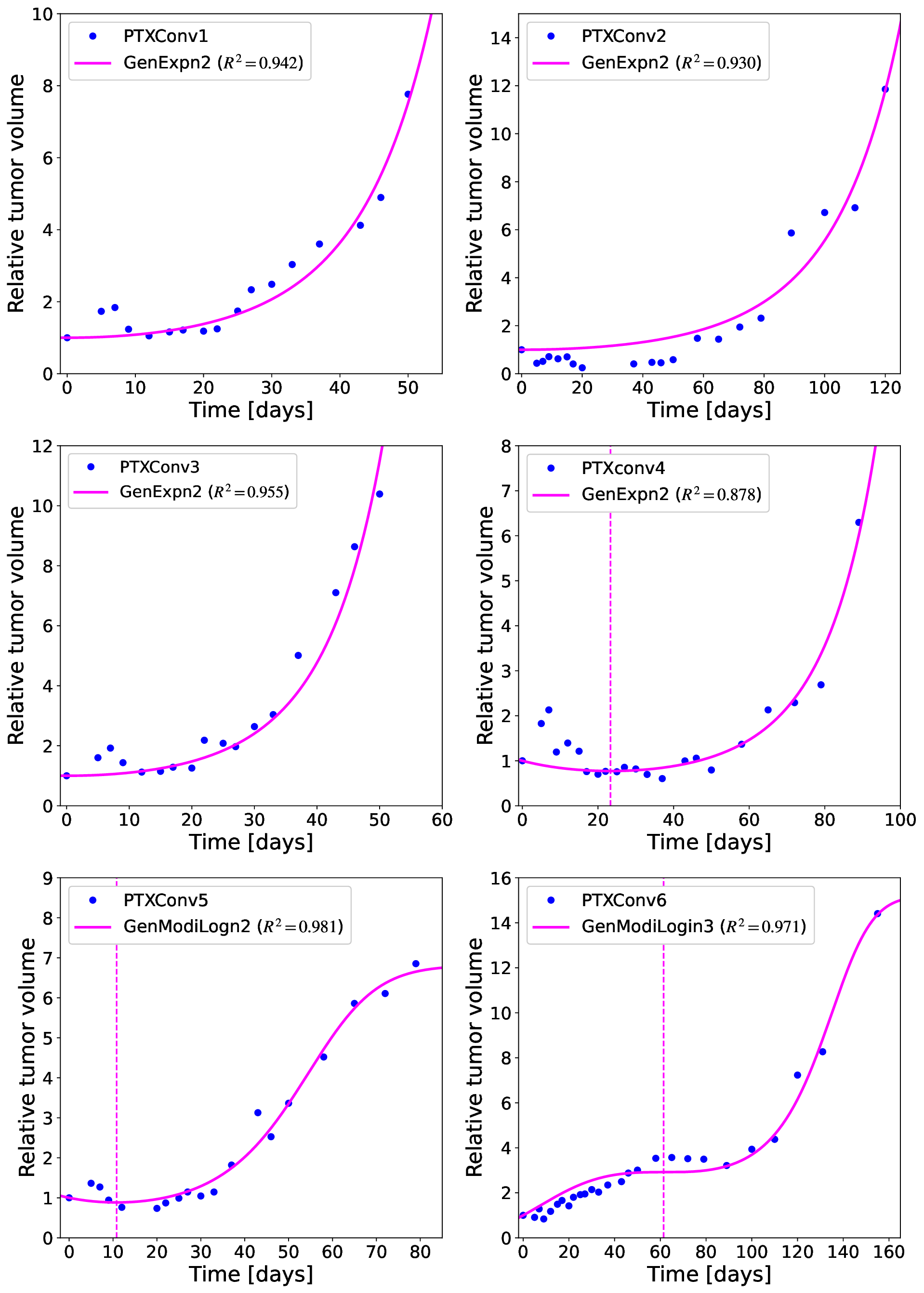}
\caption{Fits of the data sets named PTXConv in \cite{falcetta2017modeling}. The points represent the experimental values and the continuous lines represent the best fits. The dashed vertical lines in some plots mark the position of $c$. We observe that none of the models predict a vanishing volume in the far future}
\label{Fig:AllPTXConv}
\end{figure} 

\begin{figure}[H]
\centering
\includegraphics[scale=0.4]{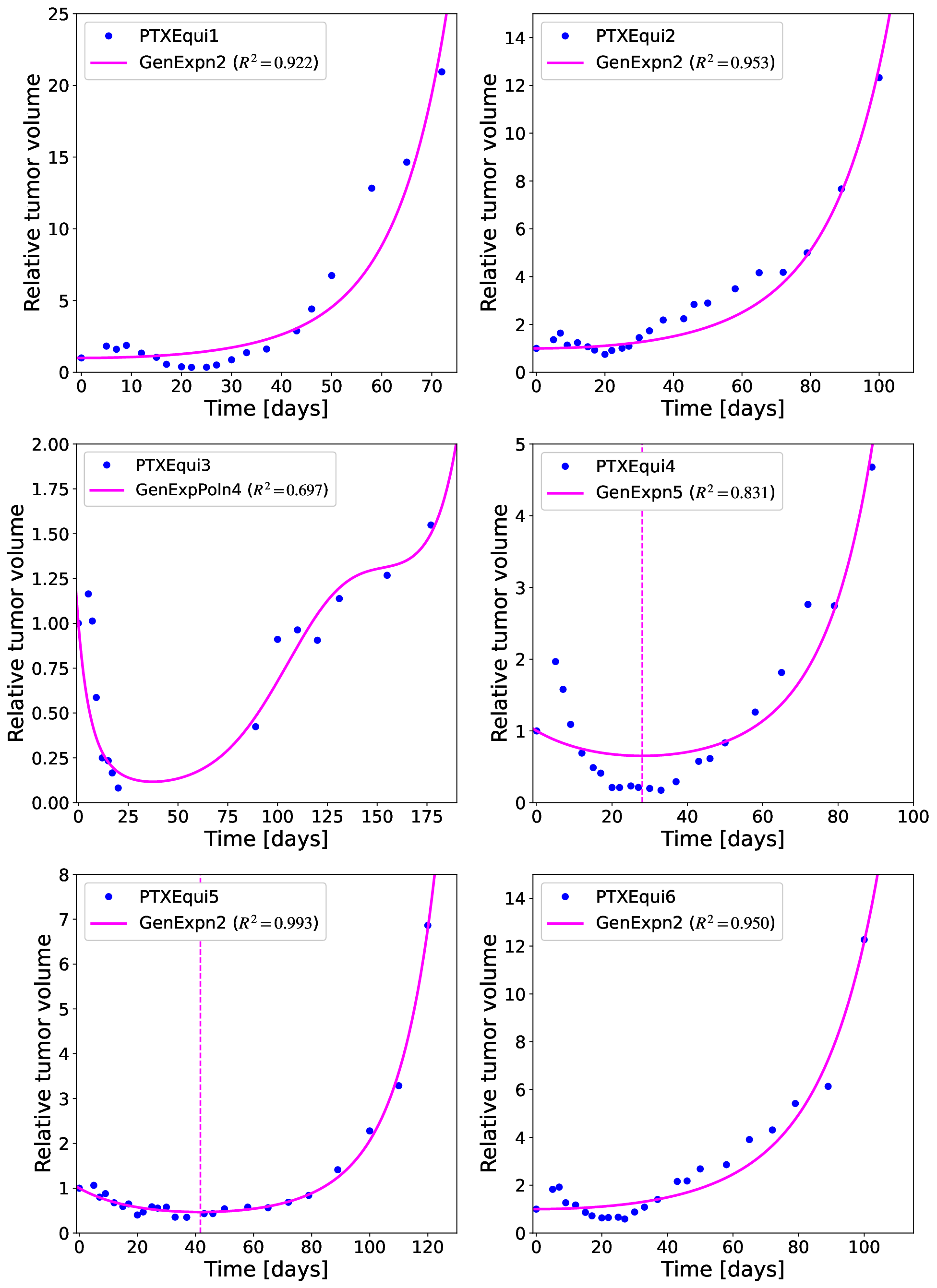}
\caption{Fits of the data sets named PTXEqui in \cite{falcetta2017modeling}. The points represent the experimental values and the continuous lines represent the best fits. The dashed vertical lines in some plots mark the position of $c$. We observe that none of the models predict a vanishing volume in the far future}
\label{Fig:AllPTXEqui}
\end{figure} 

\begin{figure}[H]
\centering
\includegraphics[scale=0.4]{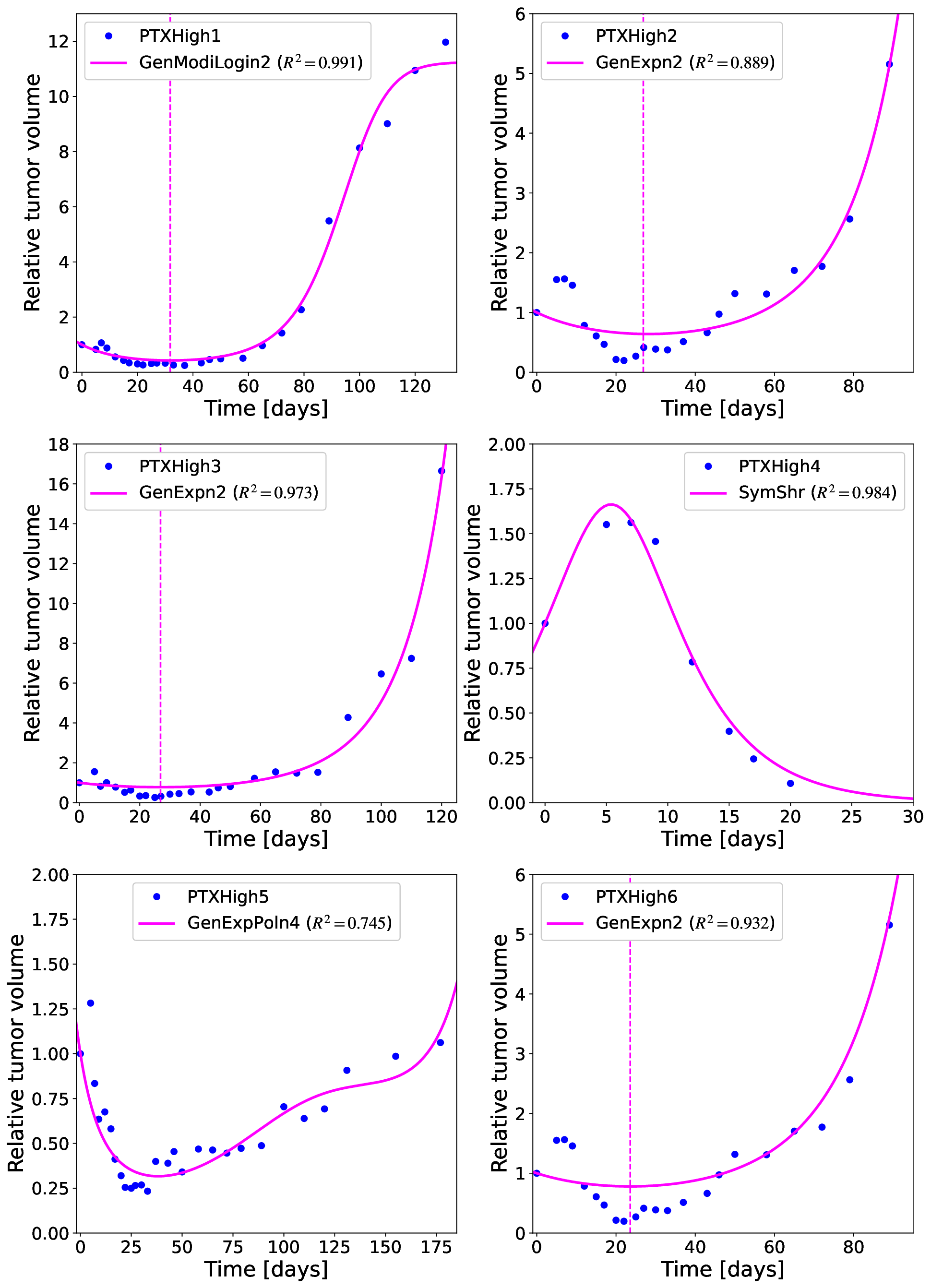}
\caption{Fits of the data sets named PTXHigh in \cite{falcetta2017modeling}. The points represent the experimental values and the continuous lines represent the best fits. The dashed vertical lines in some plots mark the position of $c$. We observe that only in PTXHigh4 there is volume shrinkage when $t\rightarrow\infty$}
\label{Fig:AllPTXhigh}
\end{figure} 

\begin{figure}[H]
\centering
\includegraphics[scale=0.4]{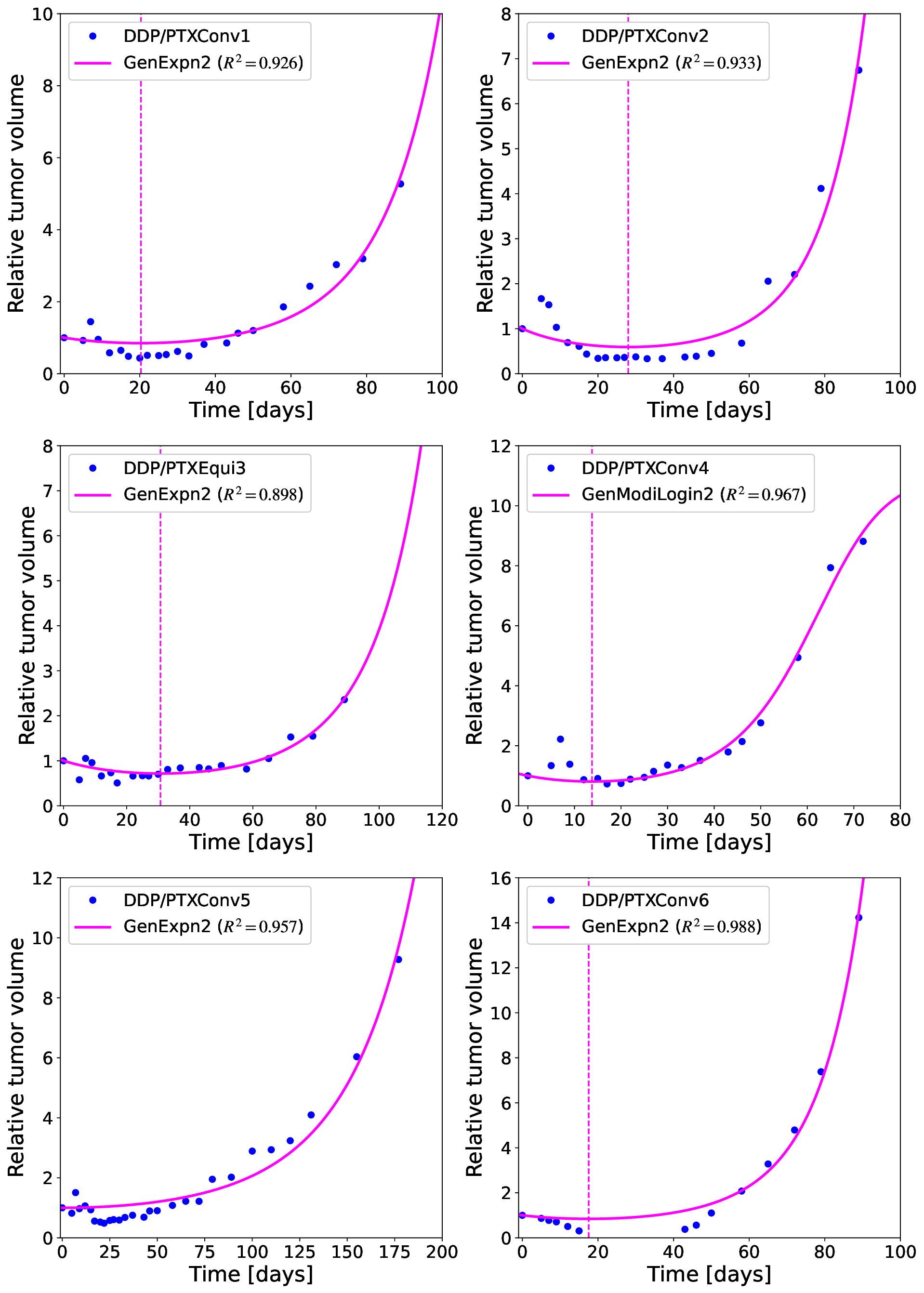}
\caption{Fits of the data sets named DDP/PTXConv in \cite{falcetta2017modeling}. The points represent the experimental values and the continuous lines represent the best fits. The dashed vertical lines in some plots mark the position of $c$. We observe that none of the models predict a vanishing volume in the far future}
\label{Fig:AllDDPPTXConv}
\end{figure} 

\begin{figure}[H]
\centering
\includegraphics[scale=0.4]{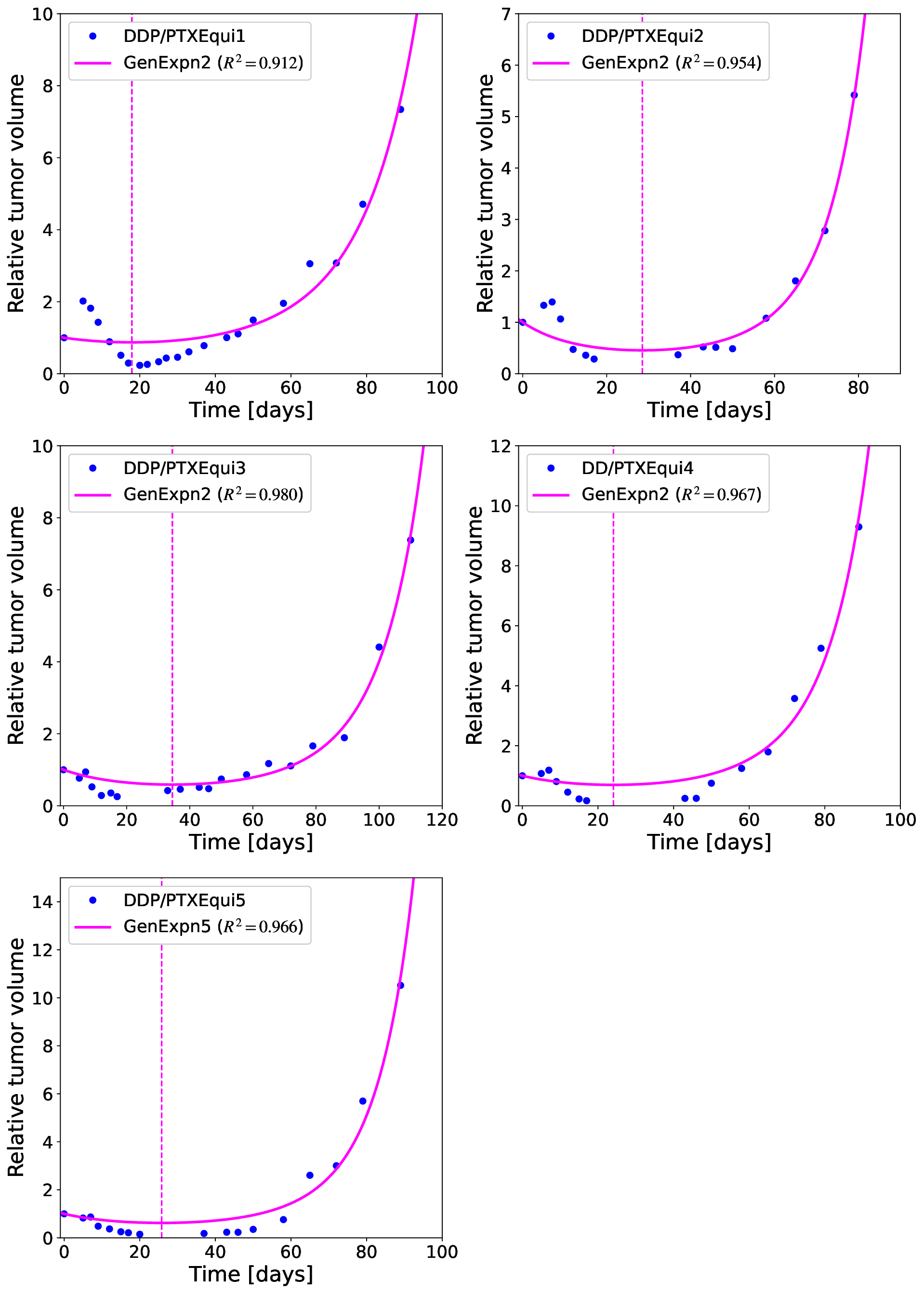}
\caption{Fits of the data sets named DDP/PTXEqui in \cite{falcetta2017modeling}. The points represent the experimental values and the continuous lines represent the best fits. The dashed vertical lines in some plots mark the position of $c$. We observe that none of the models predict a vanishing volume in the far future}
\label{Fig:AllDDPPTXEqui}
\end{figure} 

\begin{figure}[H]
\centering
\includegraphics[scale=0.4]{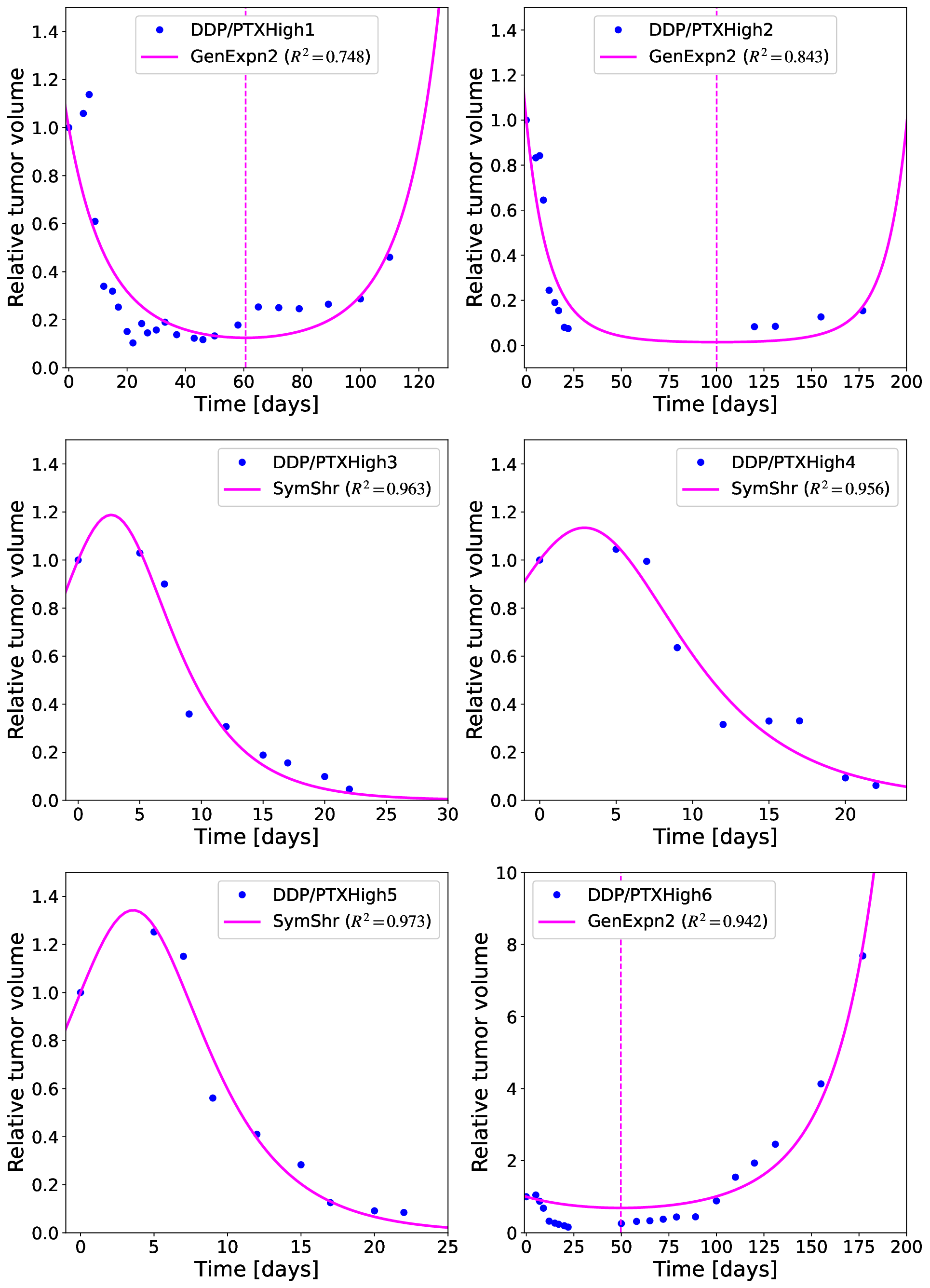}
\caption{Fits of the data sets named DDP/PTXHigh in \cite{falcetta2017modeling}. The points represent the experimental values and the continuous lines represent the best fits. The dashed vertical lines in some plots mark the position of $c$. We observe that there are 3 data sets for which volume shrinkage is predicted when $t\rightarrow\infty$}
\label{Fig:AllDDPPTXhigh}
\end{figure} 

\begin{figure}[H]
\centering
\includegraphics[scale=0.39]{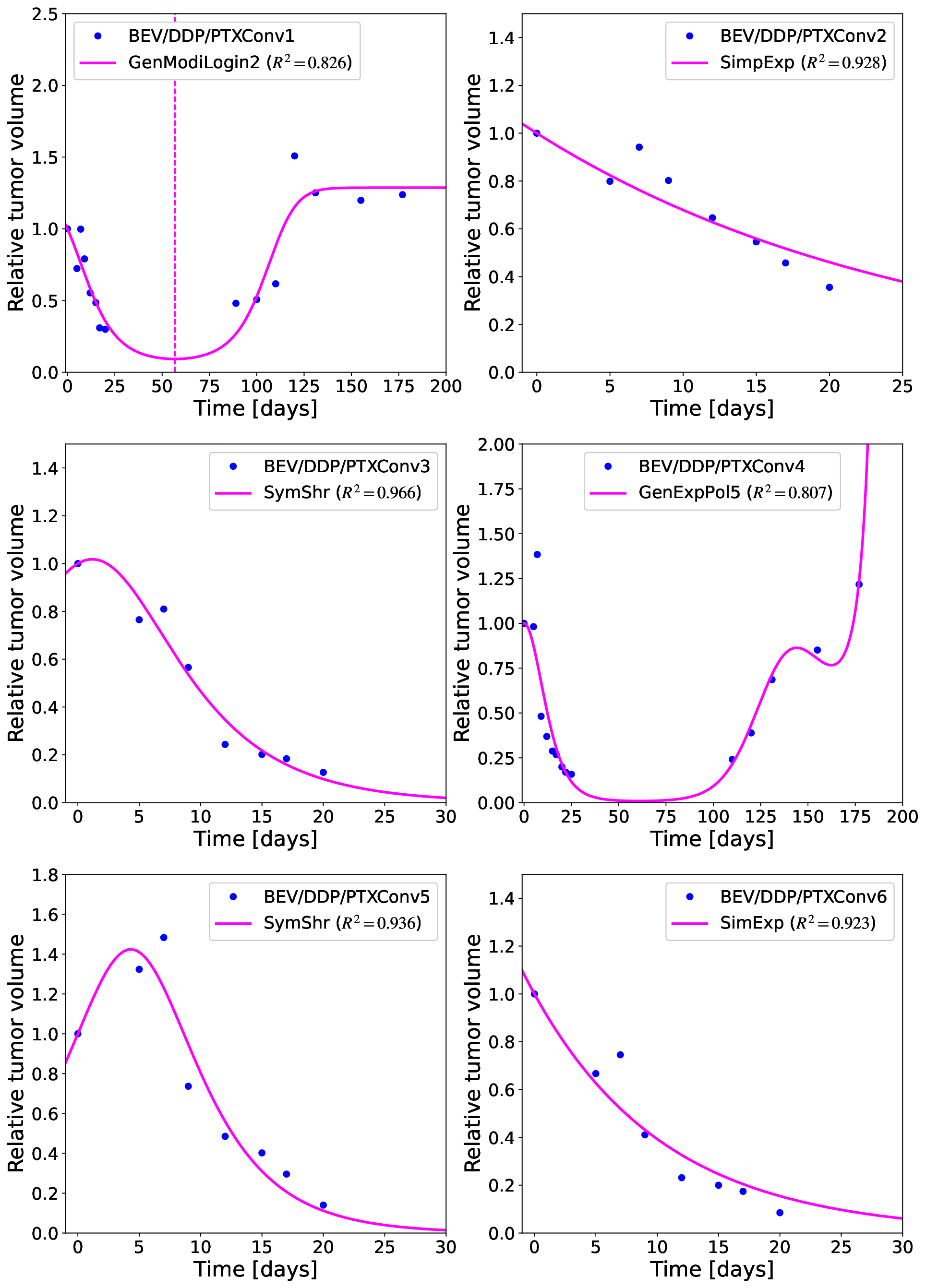}
\caption{Fits of the data sets named BEV/DDP/PTXConv in \cite{falcetta2017modeling}. The points represent the experimental values and the continuous lines represent the best fits. The dashed vertical line in the set BEV/DDP/PTXConv1 marks the position of $c$. We observe that the majority of our fits predict volume shrinkage when $t\rightarrow\infty$. On the other hand, the polynomial used for BEV/DDP/PTXConv4 does not contain the liner term}
\label{Fig:AllBEVDDPPTXConv}
\end{figure} 

\begin{figure}[H]
\centering
\includegraphics[scale=0.4]{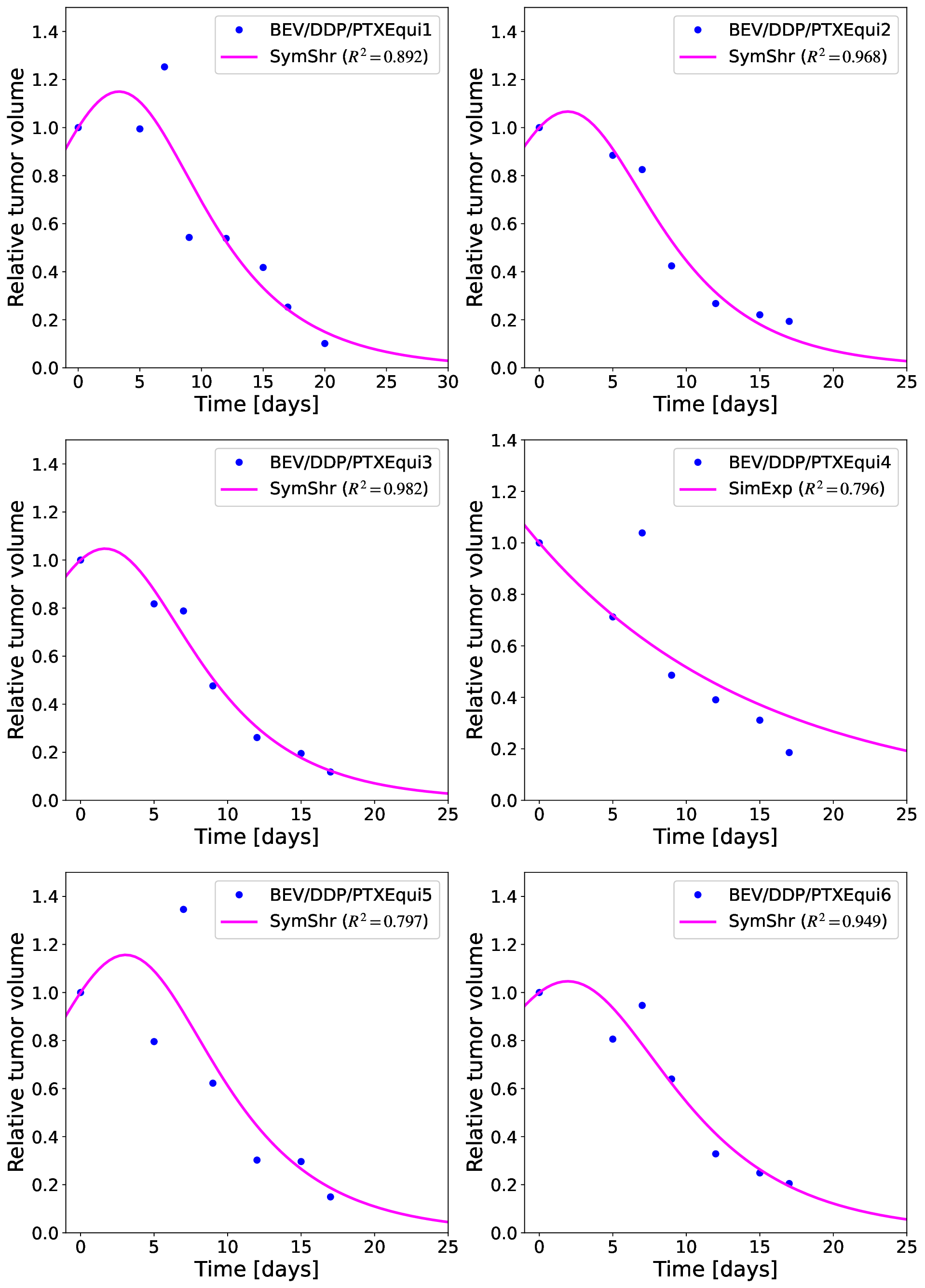}
\caption{Fits of the data sets named BEV/DDP/PTXEqui in \cite{falcetta2017modeling}. The points represent the experimental values and the continuous lines represent the best fits. We observe that all our fits predict volume shrinkage when $t\rightarrow\infty$}
\label{Fig:AllBEVDDPPTXEqui}
\end{figure} 

\begin{figure}[H]
\centering
\includegraphics[scale=0.4]{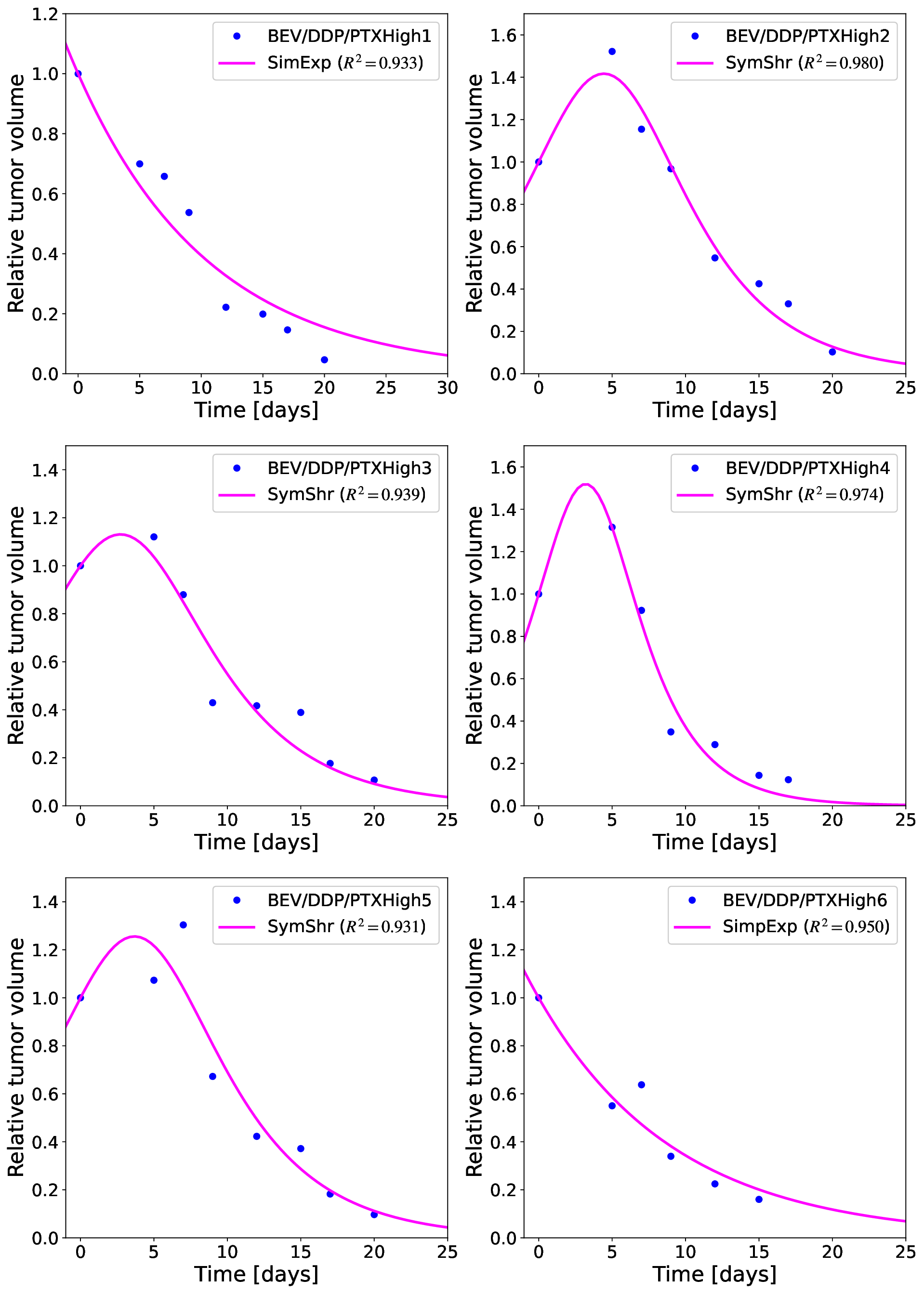}
\caption{Fits of the data sets named BEV/DDP/PTXHigh in \cite{falcetta2017modeling}. The points represent the experimental values and the continuous lines represent the best fits. We observe that all our fits predict volume shrinkage when $t\rightarrow\infty$}
\label{Fig:AllBEVDDPPTXhigh}
\end{figure} 



\bibliography{sn-bibliography}

\end{document}